\documentclass[twocolumn,preprintnumbers,amsmath,amssymb]{revtex4}
\usepackage{tabularx}
\usepackage{rotating}
\usepackage{graphicx}
\usepackage{epsfig}   
\usepackage{amsmath,amssymb,bbm}
\usepackage{anysize}
\usepackage{subfigure}

\marginsize{1in}{1in}{1in}{1in}

\newcommand{\parc}[2]{\frac{\partial #1}{\partial #2}} 
\newcommand{\units}{\, \mathrm}

\begin{document}

\title{Adaptive Resolution Molecular Dynamics Simulation:\\ Changing the
Degrees of Freedom on the Fly}

\author{Matej Praprotnik}
\altaffiliation{On leave from National Institute of Chemistry, Hajdrihova 19,
                   1000 Ljubljana, Slovenia. Electronic Mail: praprot@cmm.ki.si.}
\author{Luigi Delle Site}
\author{Kurt Kremer}
\affiliation{%
Max-Planck-Institut f\"ur Polymerforschung, Ackermannweg 10, D-55128 Mainz, Germany
}%

\date{\today}

\begin{abstract}
 We present a new adaptive resolution
technique for efficient particle-based multiscale molecular
dynamics (MD) simulations. The presented approach is tailor-made
for molecular systems where atomistic resolution is required only
in spatially localized domains whereas a lower mesoscopic level of
detail is sufficient for the rest of the system. Our method allows
an on-the-fly interchange between a given molecule's atomic and
coarse-grained level of description, enabling us to reach large
length and time scales while spatially retaining atomistic details
of the system. The new approach is tested on a model system of a
liquid of tetrahedral molecules. The simulation box is divided
into two regions: one containing only atomistically resolved
tetrahedral molecules, the other containing only one particle
coarse-grained spherical molecules. The molecules can freely move
between the two regions while changing their level of resolution
accordingly. The coarse-grained and the atomistically
resolved systems have the same statistical properties at the same
physical conditions.
\end{abstract}

\maketitle

\section{Introduction}
Many problems in complex soft matter systems are inherently
multiscale in nature, i.e., microscopic interactions are coupled
strongly to meso- and macroscopic properties. Despite the
increasing computational power and ongoing efforts to enhance the
efficiency of molecular dynamics (MD) integration
algorithms\cite{Deuflhard:1999,Minary:2004,
Praprotnik:2005,Praprotnik:2005:1,Praprotnik:2005:2,Praprotnik:2004},
all-atom MD simulations are often incapable to cover the time and
length  scales needed in order to reach relaxation in a typical
molecular system, such as a polymer solution or melt. In many
cases it is also questionable, whether the huge amount of detail
information might not even obscure the relevant structural
information. On the other hand, details of the chemistry do not
affect universal power laws but the prefactors of these power
laws, which can vary by several orders of magnitude themselves.
Thus even on the more coarse-grained level it is advisable to keep
a link to the specific chemistry under
investigation\cite{kremer:2001,kremer:2004}. In addition typical
soft matter systems can be quite inhomogeneous in a way, that
different regions within one system are sufficiently described by
more or less detail. A consistent and at the same time highly
efficient ansatz to understand modern soft matter systems (both of
synthetic as well as biological origin) has to take such
considerations into account. One first way to tackle this, was to
reduce the number of degrees of freedom by a systematic
coarse-graining, which retains only those degrees of freedom that
are relevant for the particular property of interest. Examples of
molecular systems where the coarse-graining approach has been used
with much success are fluids\cite{Ayton:2004}, lipid bilayers
\cite{Kranenburg:2003,Nielsen:2004,Marrink:2004,Chang:2005,Cook:2005},
and polymer
systems\cite{Kremer:1990,Tschop:1998,Tschop:1998:2,Everaers:2004}.

Since some specific chemical details are usually lost in the
coarse-graining procedure, much effort has been devoted recently
to the development of multiscale modeling approaches, where
different parts of the system are modeled at different levels of
detail to account for the local resolution
requirement\cite{Chun:2000,Malevanets:2000,Villa:2005,DelleSite:2002,Abrams:2003,DelleSite:2004}.
In the dual-resolution modeling approach for studying the behavior
of polymers near metal
surfaces\cite{DelleSite:2002,Abrams:2003,DelleSite:2004}, for
example, polymer chain ends that interact with the metal surface
are represented partially atomistically while the remaining parts
of the polymers, where lower resolution is adequate, are
represented as bead-spring chains. However, the switching between
different levels of resolution, i.e, the atomic and mesoscopic, is
not allowed during the course of that MD simulation, and therefore
the initial level of detail and thereby the number of degrees of
freedom in the system remain unchanged. Since the chain ends that
interact with the metal typically remain close to the surface and
they only contain a small fraction of the whole system, an
adaptive on-the-fly change of the molecules' resolution is not
strictly required for this class of systems. Another approach
reported in the literature concerns the link between quantum
mechanical and classical MD simulations. In this QM/MM approach a
small subset of the system is defined and considered as quantum
mechanical, while the rest is treated by a classical force field
simulation. Here the atoms as well as the regions of the
different regimes are fixed from the very beginning
\cite{Laio:2002}, restricting the application to rather specific
cases.

In contrast, MD simulations, in which the spatially localized
atomistic domains frequently exchange particles with the remaining
mesoscopic part of the system, would allow for much wider
applications. Then it would be possible to define certain areas or
develop criteria for certain situations, which ask for a more
detailed view, while the rest of the system can be treated on the
more coarse-grained level. It is the purpose of this paper to
present a first attempt of an MD simulation of that kind. Because
of that we here at first restrict ourselves to a most elementary
model system. Of course one also could resort to Monte Carlo
simulations, where different levels of detail are combined. This
actually would be somewhat simpler because of the purely
stochastic nature of this simulation method. Since we are
eventually aiming at molecular systems, where collective motions
are crucial, we decided to stick to the MD approach. Existing
hybrid MD methods that concurrently couple different length scales
have been developed to study solid state systems, where atomistic
MD was either combined with the finite elements
method\cite{Rafii:1998,Broughton:1999,Smirnova:1999} or it was
linked to a quantum mechanical model\cite{Csanyi:2004}. To our
knowledge, however, in pursuit of this objective no adaptive
hybrid atomistic/mesoscale particle-based MD method, which would
allow to dynamically adjust the level of detail, which means the
adjustment of the degrees of freedom in the system, has been
developed so far.

In this paper we present a novel adaptive resolution MD scheme
that combines a full atomistic description of a desired region of
the system with a mesoscale treatment of the remaining part. The
key feature of the new method is that it allows to dynamically
adapt the level of a given molecule's resolution according to its
position in the system. Hence, the number of degrees of freedom is
not a conserved quantity in our MD simulations. Furthermore, the
presented method is not restricted to couple only the atomistic
and mesoscopic levels of detail but can also be applied to systems
with mesoscopic domains that are described at different levels of
coarse-graining. Here we present a first test case, showing that
such an approach is feasible. Therefore the new approach is tested
for a simple model liquid consisting of tetrahedral molecules. The
simulation box is divided into two regions: one containing
''atomistically'' resolved tetrahedral molecules, the other
containing coarse-grained spherical molecules. Molecules are
allowed to freely move between the two regions while changing
their level of resolution accordingly. The results show that the
statistical properties of the corresponding fully atomistic system
are accurately reproduced by using the proposed hybrid scheme. In
particular, gradients in the chemical potential across the
artificial interface where the resolution changes and
corresponding spurious fluxes can be avoided. Although we applied
the new method here to a generic test system it should find
application in more realistic physical systems. There has been an
initial attempt reported in Ref. \cite{Abrams:2004}, where MD
simulation of an inhomogeneously coarse-grained system of liquid
methane has been described. There the starting point were two
established models for methane, a five site atomistic and a one
site spherical where the interaction between molecules of
different species was derived by standard Lennard-Jones mixing
rules with the hydrogens of the atomistic invisible to the
spherical molecules. In this case it turned out that an effective
flux between the two different regimes occurred. This effect is
partially due to the different equilibrium state points described
by the two models, but in any case in that approach no effective
potential between coarse-grained molecules was derived. Our
approach differs from that since here we derive such potentials in
such a way that the two regimes are in true thermodynamic
equilibrium.

The organization of the article is as follows. In section 2 the
methodology is presented, whereby in the first step a
coarse-grained model is derived and parameterized from a fully
atomistic system and then, in the second step, the atomic and
mesoscopic length scales are systematically coupled in a hybrid
atomistic/mesoscopic model. The computational details are given in
section 3. The results and discussion are presented in section 4,
followed by conclusions in section 5.

\section{Methodology}

In this section, the model systems are described and the
methodology for the adaptive multiscale MD simulations is
presented.

\subsection*{Models}

\subsubsection*{All-Atom Model}

\begin{figure}[!ht]
\centering
\subfigure[]{\includegraphics[width=7.5cm]{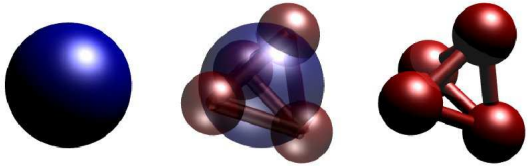}}\\
\subfigure[]{\includegraphics[width=7.5cm]{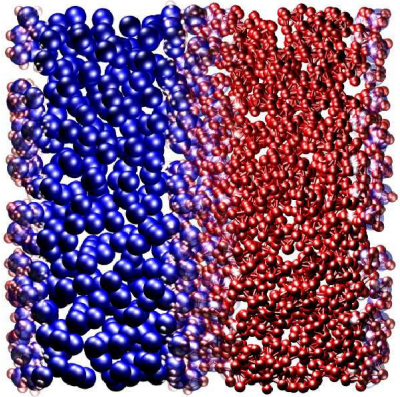}}
\caption{(a) The on-the-fly interchange between the atomic and coarse-grained
levels of description. The middle hybrid molecule is a linear
 combination of fully atomistic tetrahedral molecule with an additional center of mass
 particle representing the coarse-grained molecule. (b) Snapshot of the hybrid atomistic/mesoscopic
 model at $\rho^*=0.1$ and $T^*=1$ (LJ units). The red
molecules are the explicit atomistically resolved tetrahedral
molecules, the blue molecules are the corresponding one particle
coarse-grained molecules.}\label{Fig.1}
\end{figure}

First, we introduce our reference explicit all-atom (\emph{ex})
model. Consider a system of $n$ tetrahedral molecules consisting
of $N=4$ atoms of the same mass $m_0$ connected by anharmonic
bonds as presented in figure \ref{Fig.1} (a) (consider only the
right red molecule). 

All atoms in the system interact according to
a purely repulsive shifted $12$-$6$ Lennard-Jones potential with a
cutoff at $2^{1/6}\sigma$:
\begin{multline}
 U_{LJ}^{atom}(r_{i\alpha j\beta})=\\\left\{\begin{array}{rc}
4\varepsilon\bigl[\bigl(\frac{\sigma}{r_{i\alpha
      j\beta}}\bigr)^{12}-\bigl(\frac{\sigma}{r_{i\alpha j\beta}}\bigr)^6+\frac{1}{4}\bigr];
& r_{i\alpha j\beta}\le 2^{1/6}\sigma\\
                          0; & r_{i\alpha j\beta}> 2^{1/6}\sigma
                             \end{array}\right.\label{eq.8}
\end{multline}
where $r_{i\alpha j\beta}$ is the distance between the atom $i\alpha$ of the molecule
$\alpha$ and the atom $j\beta$ of the molecule $\beta$. We use
$\varepsilon$ as a unit of energy. All atoms have the same excluded
volume diameter $\sigma$, where $\sigma$ is the unit of length.
The neighboring atoms in a given
molecule $\alpha$ are linked via an attractive FENE potential
\begin{multline}
 U_{FENE}^{atom}(r_{i\alpha j\alpha})=\\\left\{\begin{array}{rc}
                  -\frac{1}{2}kR_0^2\ln\bigl[1-\bigl(\frac{r_{i\alpha
 j\alpha}}{R_0}\bigl)^2\bigr]; & r_{i\alpha j\alpha}\le R_0\\
                       \infty; & r_{i\alpha j\alpha}> R_0
                  \end{array}\right.\label{eq.9}
\end{multline}
with divergence length  $R_0=1.5\sigma$ and stiffness
$k=30\varepsilon/\sigma^2$, so that the average bond length is
approximately $1.0\sigma$ for $k_BT=\varepsilon$, where $T$ is the
temperature of the system and $k_B$ is Boltzmann's constant.
The functional form of these potentials and their parameters are
the same as usually employed in polymer MD simulations\cite{Kremer:1990}.

\subsubsection*{Coarse-Grained Model}

Next, we map the atomistic model to a coarse-grained (\emph{cg})
mesoscopic model. For the latter we have chosen a system composed
of $n$ one-particle molecules schematically depicted in figure
\ref{Fig.1} (a) (consider only the left blue molecule). A given
coarse-grained molecule $\alpha$ in the system has a mass
$M_\alpha=4m_0$ equal to the total mass of the explicit
tetrahedral molecule. All rotational and vibrational degrees of
freedom of atomistically resolved tetrahedral molecules are thus
removed, and the number of nonbonded interactions is strongly
decreased as well.

We shall now find an effective pair potential between
coarse-grained molecules such that the structural properties of
the underlying atomistic model are reproduced. There is usually no
unique way to coarse-grain to an effective pair potential, which
is in general temperature and density dependent
\cite{Louis:2002,Curtarolo:2002,Klapp:2004}, just as in
statistical mechanics there are different ways to perform a
renormalization group step. Here, we extract the effective pair
potential from a center-of-mass radial distribution function
(RDF$_{cm}$) of the reference atomistic model using the potential
of mean force $PMF(r)$ as
 \begin{equation}
  U^{cm}(r)\approx PMF(r)=-k_BT\log g^{cm}_{ex}(r),\label{eq.0}
\end{equation}
where $g^{cm}_{ex}(r)$ is the RDF$_{cm}$ of the all-atom system
and $U^{cm}(r)$ is the derived effective pair
potential\cite{Reith:2003}. The effective potential obtained in
this way is correct only in the limit of zero density, where the
many-body contributions vanish. For systems with nonzero densities
in principle many-body interactions would be needed just as for
the classical renormalization group theory in statistical
mechanics. In a similar spirit we here resort to the (expected)
relevant part in order to gain a significant speed up in our
simulations. Because of that we use the PMF as the initial guess
for the effective pair potential in systems with nonzero density.
Then this is further fine-tuned until the RDF$_{cm}$s and 
pressures of the reference atomistic and coarse-grained systems
match\cite{Reith:2003}. As it turns out (see the Results and
Discussion section) the effective pair potential acting between
our coarse-grained molecules is significantly softer than the pair
potential between atoms of the resolved molecules in accordance
with the results previously found in the
literature\cite{Klapp:2004,Bryant:2002}.

\subsubsection*{Transition Regime: Hybrid Atomistic/Mesoscopic Model}
Let us now introduce a hybrid explicit/coarse-grained
(\emph{ex-cg}) model. Consider a box of $n$ molecules where one
half of the box is occupied by atomistically resolved four-atom
tetrahedral molecules while the other half is filled up with the
same number of corresponding coarse-grained one-particle molecules
as schematically presented in figure \ref{Fig.1} (b). The two
domains exchange molecules which adapt their level of resolution
accordingly. To ensure that the transition between the two
different levels of description is smooth, i.e., the rotational and
vibrational degrees of freedom of a given molecule are gradually
'switched on/off' as it crosses the boundary between the atomistic
and coarse-grained domains, we also introduce an interface layer
for handshaking between atomistic and mesoscale regions. In this
regime also the rotational and vibrational velocities have to be
reintroduced in a consistent way. Due to the periodic boundary
conditions employed in our simulations there are in fact two such
layers as depicted in figure \ref{Fig.1} (b).

Since the total number of molecules $n$ in the system is a
conserved quantity in our constant temperature simulations, we
sample the phase space from the $nVT$ ensemble. However, the total
number of degrees of freedom is not constant in this model.

An alternative way compared to the similarities with
renormalization group methods, which probably describes the
situation even better, is the comparison to a first order phase
transition. The rotational and vibrational part of the free energy
then can be viewed as the latent heat at this transition.  At
equilibrium, conditions analogous to two phase coexistence,
\begin{equation}
       \mu_{ex}=\mu_{cg},~~~ p_{ex}=p_{cg},~~~ T_{ex}=T_{cg},\label{eq.1}
\end{equation}
must be automatically satisfied, where $\mu_{ex}$, $p_{ex}$,
$T_{ex}$ and $\mu_{cg}$, $p_{cg}$, $T_{cg}$ are the chemical
potentials, pressures, and temperatures of the liquid in the
atomistic and coarse-grained domains, respectively. These
conditions (\ref{eq.1}) assure that there is no net flux of
molecules between the atomistic and coarse-grained regions. To
keep this absolute requirement also then defines a central task.
This guarantees that the liquid is homogeneous across the box as
it is in the reference fully atomistic system. From a
molecular point of view, the artificial resolution boundary must
be essentially invisible, i.e., the molecules have to cross the border
without experiencing any `barrier'. Our approach to reach this
objective is presented in the proceeding subsection.

\subsection*{Adaptive Resolution Scheme}

To allow a coarse-grained molecule to find an energetically
permissible orientation  with respect to its neighboring
mo\-le\-cu\-les (when it leaves the coarse-grained domain and is
remapped into the atomistically resolved four-atom tetrahedral
molecule) we introduce an interface layer between the atomistic
and coarse-grained regions, which contains 'hybrid' molecules as
presented in figure \ref{Fig.1}. Each hybrid molecule
schematically shown in figure \ref{Fig.1}(a) (consider the middle
molecule) is composed of a tetrahedral molecule with an additional
massless center-of-mass particle serving as an interaction site.
This is similar to the flexible TIP4P water
model\cite{Lawrence:2003} where apart from the interaction sites
on the three atoms of a water molecule an additional interaction
site is introduced along the symmetry axis between the hydrogen
and oxygen atoms.

Thus, each time a coarse-grained molecule $\alpha$ leaves that
domain and enters the interface layer, it is remapped first into a
hybrid molecule with the same center-of-mass position and a random
orientation in space, where the relative positions of the
tetrahedral atoms are taken from a molecular configuration
corresponding to a randomly chosen molecule from the atomistic
regime. Each of the four explicit tetrahedral atoms in the hybrid
molecule gains at this remapping a velocity equal to the velocity
of the corresponding coarse-grained molecule to maintain the
linear momentum of the molecule. In addition, the tetrahedral
atoms are also assigned rotational/vibrational velocities
corresponding to atoms of a random molecule from the atomistic
region, where we subtract the total linear momentum of the latter
molecule. In this way we ensure that the kinetic energy is
distributed among all degrees of freedom according to the
equipartition principle as $k_BT/2$ of average kinetic energy per
quadratic degree of freedom while retaining the linear momentum of
the coarse-grained molecule. The center-of-mass interaction site
moves obeying the constraints:
\begin{eqnarray}
{\bf R}_\alpha&=&\frac{\sum_{i\alpha} m_{i\alpha}{\bf
    r}_{i\alpha}}{M_\alpha},\label{eq.5}\\
 {\bf V}_\alpha&=&\frac{\sum_{i\alpha} m_{i\alpha}{\bf v}_{i\alpha}}{M_\alpha},\label{eq.6}
\end{eqnarray}
where ${\bf R}_\alpha$ is a center of mass of the molecule $\alpha$,
${\bf r}_{i\alpha}$ is the position vector of the explicit
tetrahedral atom $i\alpha$ in the molecule $\alpha$, ${\bf
V}_\alpha$ is the center-of-mass velocity of the molecule $\alpha$,
${\bf v}_{i\alpha}$ is the velocity of the explicit tetrahedral
atom $i\alpha$, and $M_\alpha=\sum_{i\alpha}m_{i\alpha}$ is the
total mass of the molecule $\alpha$. In our case,
$m_{i\alpha}=m_0$ and $M_\alpha=4m_0$ for all $i\alpha=1,\dots,4$
and $\alpha=1,\dots, n$. Each time a hybrid molecule crosses the
boundary into atomistic regime it is remapped into a four-particle
tetrahedral molecule with the four tetrahedral atoms retaining
their current velocities and positions. In this model also the
explicit tetrahedral molecules have the center-of-mass interaction
sites, but only for the interactions with the hybrid and
coarse-grained molecules. Of course, deep in the atomistic region,
where the atomistically resolved molecules do not interact
anymore with the hybrid and coarse-grained molecules the
center-of-mass interaction site can be omitted.  Every time a
four-particle tetrahedral molecule leaves the atomistic region and
enters into the transition regime it is mapped into a hybrid
molecule with the four tetrahedral atoms retaining instantaneous
velocities and positions with the center-of-mass interaction site
moving according to Eqs. (\ref{eq.5}) and (\ref{eq.6}). Similarly,
as a hybrid molecule crosses a boundary to the coarse-grained
region it is mapped into a coarse-grained molecule with a velocity
equal to the center-of-mass velocity of the hybrid molecule given by Eq.
(\ref{eq.6}).

To couple the atomic and mesoscopic length scales we define in the
spirit of thermodynamic perturbation
approach\cite{Zwanzig:1954,Leach:2001} the total intermolecular
force acting between centers of mass of molecules $\alpha$ and
$\beta$ as
\begin{multline}
 {\bf F}_{\alpha\beta}=\\w(X_\alpha)w(X_\beta){\bf
 F}_{\alpha\beta}^{atom}+[1-w(X_\alpha)w(X_\beta)]{\bf
 F}_{\alpha\beta}^{cm},\label{eq.4}
\end{multline}
where
\begin{equation}
{\bf F}_{\alpha\beta}^{atom}=\sum_{i\alpha, j\beta}{\bf F}_{i\alpha\label{eq.4a}
 j\beta}^{atom}
\end{equation}
is the sum of all pair atom interactions between explicit
tetrahedral atoms of the molecule $\alpha$ and explicit tetrahedral
atoms of the molecule $\beta$ and
\begin{eqnarray}
 {\bf F}_{i\alpha j\beta}^{atom}&=&- \parc{U^{atom}}{{\bf r}_{i\alpha j\beta}},\\
 {\bf F}_{\alpha\beta}^{cm}&=&-\parc{U^{cm}}{{\bf R}_{\alpha\beta}}.
\end{eqnarray}
The vector ${\bf r}_{i\alpha j\beta}={\bf r}_{i\alpha}-{\bf
r}_{j\beta}$ is the relative position vector of atoms $i\alpha$
and $j\beta$, ${\bf R}_{\alpha\beta}={\bf R}_{\alpha}-{\bf
R}_{\beta}$ is the relative position vector of the centers of mass
of the molecules $\alpha$ and $\beta$, $X_\alpha$ and $X_\beta$ are
the $x$ center-of-mass coordinates of the molecules $\alpha$ and
$\beta$, respectively, and $w$ is the weighting function that
determines the 'identity' of a given molecule. The weighting
function $w\in [0,1]$ is defined in such a way that values $0<w<1$
correspond to a hybrid molecule with extreme cases $w=1$ and $w=0$
corresponding to a four-atom tetrahedral molecule and one-particle
coarse-grained molecule, respectively. Hence, as soon as one of
the two interacting molecules $\alpha$ and $\beta$ is a
coarse-grained molecule with no explicit tetrahedral atoms
$w(X_\alpha)w(X_\beta)=0$ and ${\bf F}_{\alpha\beta}={\bf
F}_{\alpha\beta}^{cm}$. 

\begin{figure}[!ht]
\centering
\mbox{\includegraphics[width=7.5cm, angle=0]{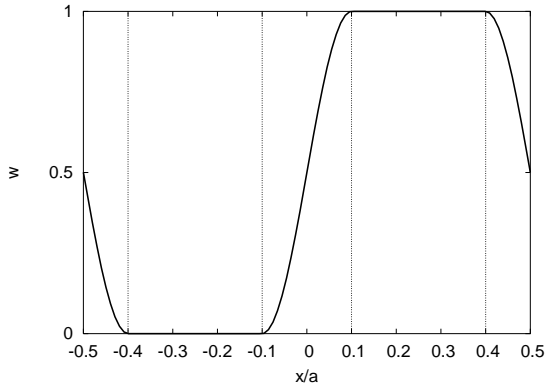}}
         \caption{ The weighting function $w(x)\in [0,1]$ defined by
  Eq. (\ref{eq.7}). The values $w=1$ and $w=0$ correspond to the atomistic
  and coarse-grained regions of the hybrid atomistic/mesoscopic system
  with the box length $a$, respectively, while the values $0<w<1$ correspond to
  the interface layer. Shown is the example where the half-width $d$ of
  the interface layer is $d=a/10$. The vertical lines denote the
  boundaries of the interface layers.}\label{Fig.2}
\end{figure}

We propose the following functional form of the weighting function
$w$:
\begin{multline}
 w(x)=\\\left\{\begin{array}{rc}
                                                   1; & d<x\le \frac{a}{2}-d\\
                                                   0; & -\frac{a}{2}+d\le x<-d\\
                         \sin^2[\frac{\pi}{4d}(x+d)]; & -d\le x\le d\\
             \cos^2[\frac{\pi}{4d}(x-\frac{a}{2}+d)]; & \frac{a}{2}-d<x\le\frac{a}{2}\\
             \cos^2[\frac{\pi}{4d}(x+\frac{a}{2}+d)]; & -\frac{a}{2}\le x<-\frac{a}{2}+d
            \end{array}\right.\label{eq.7}
\end{multline}
where $a$ is the box length and $d$ the half-width of the interface
layer. The weighting function $w$ is shown in figure \ref{Fig.2}. Our
choice, which takes into account the periodic boundary conditions,
 is a particularly simple way to ensure an interpolation between
$w=0$ and $w=1$ that is monotonic, continuous, differentiable and
has zero slope at the boundaries to the atomistic and
coarse-grained regions. Apart from these requirements we consider
the precise functional form as immaterial.

Exploiting the analogy with the quantum mechanical mixed state
description\cite{Schwabl:1995}, one can consider a hybrid molecule
in the interface layer as a normalized linear combination of a
four-atom tetrahedral molecule and a corresponding one-particle
coarse-grained molecule. As a given molecule moves from the
coarse-grained boundary of the interface layer to the atomistic
boundary, $w$ is gradually changed from $0$ to $1$ and a
coarse-grained molecule with only $3$ translational degrees of
freedom gradually turns into an atomistically resolved molecule
with additional $3N-3=9$ rotational and vibrational degrees of
freedom and a defined spatial orientation. The continuous, i.e.,
not instantaneous, 'identity' transition is required since a
hybrid molecule is given a random orientation at the
coarse-grained boundary, and there can consequently be overlaps of
its tetrahedral atoms with the atoms of the neighboring molecules.
Since at this boundary $w=0$ and the repulsive potential
$U^{atom}$ given in Eq. (\ref{eq.8}) is capped (see the
Computational Details section), the forces acting on these atoms
cannot diverge, however. While moving towards the atomistic region
the hybrid molecule slowly adapts its orientation via the
gradually increasing atomistic interactions with the neighboring
molecules. Likewise, as presented in figure \ref{Fig.1} (a), as a
given molecule moves from the atomistic boundary of the interface
layer to the coarse-grained boundary, $w$ is continously changing
from $1$ to $0$, the fully atomistically resolved molecule
gradually turns into the one particle coarse-grained molecule
while omitting all rotational and vibrational degrees of freedom
and orientation.

To elucidate the definition of the force calculation in our
model we rewrite Eq. (\ref{eq.4}) as
\begin{multline}
{\bf F}_{\alpha\beta}=\\
w(X_\alpha)w(X_\beta){\bf F}_{\alpha \beta}^{atom}+[1-w(X_\alpha)][1-w(X_\beta)]{\bf
  F}_{\alpha\beta}^{cm}\\+[1-w(X_\alpha)]w(X_\beta){\bf
  F}_{\alpha\beta}^{cm}+w(X_\alpha)[1-w(X_\beta)]{\bf F}_{\alpha\beta}^{cm}.\label{eq.3}
\end{multline}
From Eqs. (\ref{eq.4a}) and (\ref{eq.3}) one can then deduce that
the pair force in Eq. (\ref{eq.4}) is defined in such a way that
two given atoms $i\alpha$ and $j\beta$ in given explicit molecules
$\alpha$ and $\beta$ ($w(X_\alpha)=1$ and $w(X_\beta)=1$) interact
via the atomistic potential defined by Eqs. (\ref{eq.8}) and
(\ref{eq.9}) while two coarse-grained molecules ($w(X_\alpha)=0$
and $w(X_\beta)=0$) interact via the corresponding effective pair
potential $U^{cm}$. Furthermore, the coarse-grained molecules
'see' the fully atomistically resolved molecules as coarse-grained
molecules. Hence, their intermolecular interaction is defined by
the effective pair potential $U^{cm}$. To ensure that the
center-of-mass dynamics governed by Eqs. (\ref{eq.5}) and
(\ref{eq.6}) of a given explicit or hybrid molecule $\alpha$ is
correct, the total intermolecular force ${\bf
F}_{\alpha\beta}^{cm}$ between the  atomistically resolved
molecule $\alpha$ and a coarse-grained molecule $\beta$ is
distributed among the explicit atoms of the molecule $\alpha$ as
\begin{equation}
 {\bf F}_{i\alpha\beta}=\frac{m_{i\alpha}}{\sum_{i\alpha}m_{i\alpha}}{\bf F}_{\alpha\beta}^{cm},\label{eq.4b}
\end{equation}
where  ${\bf F}_{i\alpha\beta}$ is the force imposed on the
explicit tetrahedral atom $i\alpha$ by the coarse-grained molecule
$\beta$. The explicit tetrahedral atoms in a given hybrid molecule
interact with other explicit atoms in neighboring explicit and
hybrid molecules through atomistic forces, while the then massless
center-of-mass particle serves as an effective potential
interaction site. The total force on a hybrid molecule is then
according to Eq. (\ref{eq.4}) a normalized linear combination of
atomistic and effective pair forces. 
Using
\begin{eqnarray}
  {\bf F}_{i\alpha j\beta}^{atom}&=&-{\bf F}_{j\beta i\alpha}^{atom},\\
  {\bf F}_{\alpha\beta}^{cm}&=&-{\bf F}_{\beta\alpha}^{cm}
\end{eqnarray}
we obtain from Eq. (\ref{eq.4})
\begin{equation}
{\bf F}_{\alpha \beta}=-{\bf F}_{\beta \alpha}.
\end{equation}
The force definition in Eq. (\ref{eq.4}) hence satisfies Newton's Third
Law.

Recall that the effective intermolecular potential is determined in such a way that the 
equations of state for the \emph{ex} and \emph{cg} models match around the state
point considered. Therefore, following the scheme, as given by Eq.
(\ref{eq.4}), Eq. (\ref{eq.1}) is implicitly
satisfied, due to which spurious fluxes are avoided at the boundary between
the atomistic and coarse-grained regimes.

To summarize, the new adaptive resolution scheme for the hybrid
atomistic/mesoscale MD simulations is a two-stage procedure:
\begin{enumerate}
 \item{Derive the effective
pair potential $U^{cm}$ between coarse-grained molecules on the basis of the
reference all-atom system.}
\item{Introduce the interface layer containing the hybrid molecules that have
  additional interaction sites positioned at their centers of mass. 
  Define a weighting function $w$ by Eq. (\ref{eq.7}) and
  use Eq. (\ref{eq.4}) for the definition of the intermolecular pair forces.
  Allow molecules to adapt their level of resolution according to
  their position in the system as explained in the second paragraph of
  this subsection.}
\end{enumerate}
Finally, since the switching of the resolution can be considered as a first
order phase transition, the adaptive resolution scheme must
necessarily be employed in combination with a thermostat. 
Because the latent heat is generated in the
transition regime it is important to use a thermostat, which 
couples locally to the particle motion, e.g., Langevin or Dissipative Particle Dynamics (DPD) thermostats\cite{Soddemann:2003}.

\section{Computational Details}

\subsection*{Temperature Calculation}

In order to treat all $n$ molecules equally regardless of the
level of detail, we define, using the equipartition principle, the
temperature of a system only from translational degrees of
freedom. The 'translational' temperature
\begin{equation}
 T=\frac{2E_k^{cm}}{3nk_B}=\frac{1}{3nk_B}\sum_\alpha M_\alpha{\bf V}^2_\alpha,\label{eq.12}
\end{equation}
when averaged gives the temperature of the system
\cite{Allen:1987}. Here $E_k^{cm}$ is the total translational
kinetic energy of the system.  We also checked the ``particle''
temperature in the explicit atomistic region. As to be expected we
find the same temperature.

\subsection*{Pressure Calculation}

For the same reason also the pressure calculation is based on the
molecular instead of atomic interactions:
\begin{equation}
 p=\frac{1}{V}\biggl[nk_BT-\frac{1}{3}\sum_{\alpha<\beta}{\bf R}_{\alpha\beta}\cdot{\bf F}_{\alpha\beta}\biggr],\label{eq.12a}
\end{equation}
where ${\bf F}_{\alpha\beta}$ is given by Eqs.
(\ref{eq.4})\cite{Berendsen:1984, Allen:1987}. Moreover, using Eq.
(\ref{eq.12a}) for pressure evaluation has two additional
advantages compared to the pressure calculation based on atomic
interactions:  first, Eq. (\ref{eq.12a}) is also valid in the case
that forces on atoms involve internal non-pairwise-additive
contributions, and second, even if all interactions are pairwise
additive (as in our case), the pressure calculation based on
atomic interactions introduces additional fluctuations in the
pressure\cite{Berendsen:1984}.

\subsection*{Multiscale Simulation Details}

For computational convenience during the proof-of-principle stage
we replace our \emph{ex-cg} model with a model in which the
whole box contains exclusively hybrid molecules with four explicit
atoms and a center-of-mass interaction site. This has some
technical advantages during the tests of the method. The true
level of detail of the molecules is then determined from the value
of $w(X_{\alpha})$. For later large scale, however, of course the
original \emph{ex-cg} model composed of the explicit,
coarse-grained, and hybrid molecules will be employed. 
Using Eq. (\ref{eq.7}) in
the replacement model for distinguishing between explicit,
coarse-grained, and hybrid molecules of our original \emph{ex-cg}
model we can capture all definitions of pair forces for all
classes of particles in our system in a single expression as
\begin{multline}
 {\bf F}_{i\alpha j\beta}=\\w(X_\alpha)w(X_\beta){\bf F}_{i\alpha
 j\beta}^{atom}+[1-w(X_\alpha)w(X_\beta)]\delta_{\alpha,\beta}{\bf F}_{i\alpha
 j\beta}^{atom}\\+[1-w(X_\alpha)w(X_\beta)]\frac{m_{i\alpha}m_{j\beta}}{\sum_{i\alpha}m_{i\alpha}\sum_{j\beta}m_{j\beta}}{\bf F}_{\alpha\beta}^{cm},\label{eq.2}
\end{multline}
where ${\bf F}_{i\alpha j\beta}$ is the total pair force between
the explicit atom $i\alpha$ of the molecule $\alpha$ and the
explicit atom $j\beta$ of the molecule $\beta$ and
$\delta_{\alpha,\beta}$ is the Kronecker symbol. Summing for
$\alpha\ne \beta$
\begin{equation}
{\bf F}_{\alpha\beta}=\sum_{i\alpha, j\beta}{\bf F}_{i\alpha j\beta}
\end{equation}
we regain the total force between molecules $\alpha$ and $\beta$
given in Eq. (\ref{eq.4}). From Eqs. (\ref{eq.5}), (\ref{eq.6}),
and (\ref{eq.2}) follows that in the replacement model with only
hybrid molecules a hybrid molecule experiences only translational
kicks from other molecules in the coarse-grained region ($w=0$)
and hence its center of mass moves exactly as the respective
one-particle coarse-grained molecule in the original \emph{ex-cg}
model. Similarly, in the explicit region ($w=1$) a hybrid molecule
experiences only atomistic forces and hence its explicit atoms
move exactly as the explicit atoms in the respective tetrahedral molecule in
the original model. Therefore, the model containing only hybrid
molecules interacting via the pair force defined by Eq.
(\ref{eq.2}) together with the applied Langevin
thermostat\cite{Kremer:1990} acting on each particle in the system
(to assure that the atom velocities are thermalized in accordance
with the equipartition principle) exactly mimics the original
\emph{ex-cg} model in which the temperature would also be held
constant by the Langevin thermostat. From the methodology
development point of view, these two models are therefore
identical.

This yields the Langevin equation of motion
\begin{equation}
 m_i\frac{d^2{\bf r_i}}{dt^2}={\bf F}_i-m_i\Gamma\frac{d{\bf
 r}_i}{dt}+{\bf W}_i(t),
\end{equation}
where $m_i$ is the mass of particle $i$, ${\bf F}_i$ is the total
force acting on the respective particle equal to the sum of pair
interactions given by Eq. (\ref{eq.2}), $\Gamma$ is the friction
constant, and ${\bf W}$ is the random force of a heat
bath\cite{Mann:2004}. We sample the random force from a uniform
distribution, since it has been shown that there is no advantage
of using Gaussian noise for the Langevin
thermostat\cite{Duenweg:1991}.

The value of the friction constant used in our simulations is
$\Gamma=0.5\tau^{-1}$ where
$\tau=(\varepsilon/m_0\sigma^2)^{-1/2}$. The equations of motion
are integrated for each particle of the system using the velocity
Verlet algorithm with a $0.005\tau$ time step. Here, again only
for the purpose of testing the method, we use only one time step
in the whole system. In the coarse-grained regime actually a
significantly larger time step could be used. Therefore ultimately
one would like to introduce a multiple time step algorithm.
Simulations are performed at temperature $T=\varepsilon/k_B$ and
number density $\rho=n/V=0.1/\sigma^3$. Here $n=5001$ is the
number of molecules in the system, which can be either the
explicit, coarse-grained or hybrid. If we roughly estimate the
excluded volume diameter of the coarse-grained molecule
$\sigma_{CG}$ as the distance, where the repulsive effective pair
potential between the coarse-grained molecules in our simulations
equals $k_BT$, i.e., $\sigma_{CG}\approx 1.7\sigma$, then the
number density
$\rho=0.1/\sigma^3=0.1(\sigma_{CG}/\sigma)^3/\sigma_{CG}^3\approx
0.5/\sigma_{CG}^3$ corresponds to a medium dense liquid, which is
due to the soft effective repulsive interactions rather weakly
correlated\cite{Kremer:2005}.
Periodic boundary conditions and the minimum image
convention\cite{Allen:1987} are employed. The interaction range in
the system is given by the range of the effective pair potential
between molecules  and the geometry of tetrahedral molecules, i.e.,
the most outer atoms of two tetrahedral molecules with centers of
mass slightly less than $2.31\sigma$ apart still experience the
effective potential contribution in Eq. (\ref{eq.2}). Hence, the
actual interaction range in the system is approximately
$3.5\sigma$.

All molecules are initially randomly placed in a cubic box of size
$a=36.845\sigma$. To remove the overlaps between them, a $50\tau$
long warm-up run is performed during which the repulsive
interparticle potential is capped (see capped interactions in Ref.
\cite{Espresso:2005}). Thus, at all interparticle distances,
which would lead to larger forces between particles than a
prescribed maximal force, the forces defined by the original
repulsive pair potential are replaced by repulsive central forces
of the maximal force magnitude. The latter is gradually increased
from $20\varepsilon/\sigma$ to $110\varepsilon/\sigma$ during this
warm-up phase. Afterwards an additional $250\tau$ equilibration
run is carried out where we set the maximal force magnitude to
$10^{9}\varepsilon/\sigma$, which corresponds to interparticle
distance of $0.27\sigma$. The chosen maximal force magnitude value
is so high that it has no effect on the dynamics of molecules in
the atomistic region because there atoms never come this close
together. Therefore, by this force capping we only prevent
possible force singularities that could emerge due to overlaps
with the neighboring molecules when a given molecule enters the
interface layer from the coarse-grained side as explained in the
previous section. Production runs with the
$10^{9}\varepsilon/\sigma$ force capping are then performed for
$7500\tau$, storing configurations at each $5\tau$ time interval
for analysis. We performed all our MD simulations using the
ESPResSo package\cite{Espresso:2005}, developed at our institute.

The following reduced Lennard-Jones units\cite{Allen:1987} are
used throughout:
 $m^*=m/m_0$, $r^*=r/\sigma$, $V^*=V/\sigma^3$,
 $T^*=k_BT/\varepsilon$, $U^*=U/\varepsilon$, $p^*=p\sigma^3/\varepsilon$,
 $\rho^*=n/V^*$, $t^*=t/\tau$, $D^*=D\sqrt{m_0/\varepsilon}/\sigma$,
where $D$ is the self-diffusion constant. Note that in our
simulations all atoms have a mass $m^*=1$ while all molecules have
a mass $M^*=4$.

\section{Results and Discussion}

\subsection*{Determination of the Effective Potential}

\begin{figure}[!ht]
\centering
\includegraphics[width=7.5cm, angle=0]{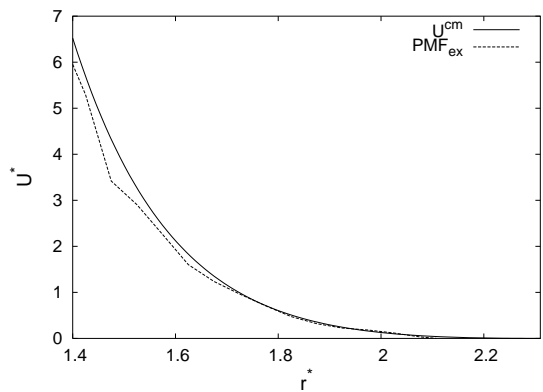}
\caption{ The effective pair potential ${U^{cm}}^*$, Eq.
(\ref{eq.11}), between the coarse-grained molecules, where the
potential of mean force PMF$_{ex}$ of the explicit system at
$\rho^*=0.0025$ and $T^*=1$ was used as the initial guess. The
presented function ${U^{cm}}^*$ was determined in such a way that
the RDF$_{cm}$s of the explicit (\emph{ex}) and coarse-grained (\emph{cg})
systems match at the $\rho^*=0.1$.}\label{Fig.3}
\end{figure}

We have determined the effective nonbonded pair potential
${U^{cm}}^*$ between coarse-grained molecules illustrated in
figure \ref{Fig.3} by using the potential of mean force
(PMF$_{ex}$) of the \emph{ex} system, Eq. (\ref{eq.0}), at very
low number density $\rho^*=0.0025$ as the initial guess. Then we
further adjusted it to obtain the adequate agreement between
RDF$_{cm}$s of the \emph{ex} and \emph{cg} systems at the
$\rho^*=0.1$.
We have parameterized the effective potential
$U^{cm}$ with the Morse potential
\begin{equation}
 {U^{cm}}^*(r^*)=\gamma^*\{1-\exp[-\kappa^*(r^*-r^*_0)]\}^2\label{eq.11}
\end{equation}
with parameters $\gamma^*=\gamma/\varepsilon=0.105$, $\kappa^*=\kappa\sigma=2.4$, and $r^*_0=r_0/\sigma=2.31$.
As one can see from figure \ref{Fig.3}, the obtained effective
potential is softer than the underlying repulsive interatomic interaction
potential given by Eqs. (\ref{eq.8}) and (\ref{eq.9}) since it varies
more slowly with the interparticle distance. This is a general feature
of effective potentials for polyatomic molecular systems\cite{Klapp:2004}.

\begin{figure}[!ht]
\centering
\includegraphics[height=10cm,width=7.5cm,angle=0]{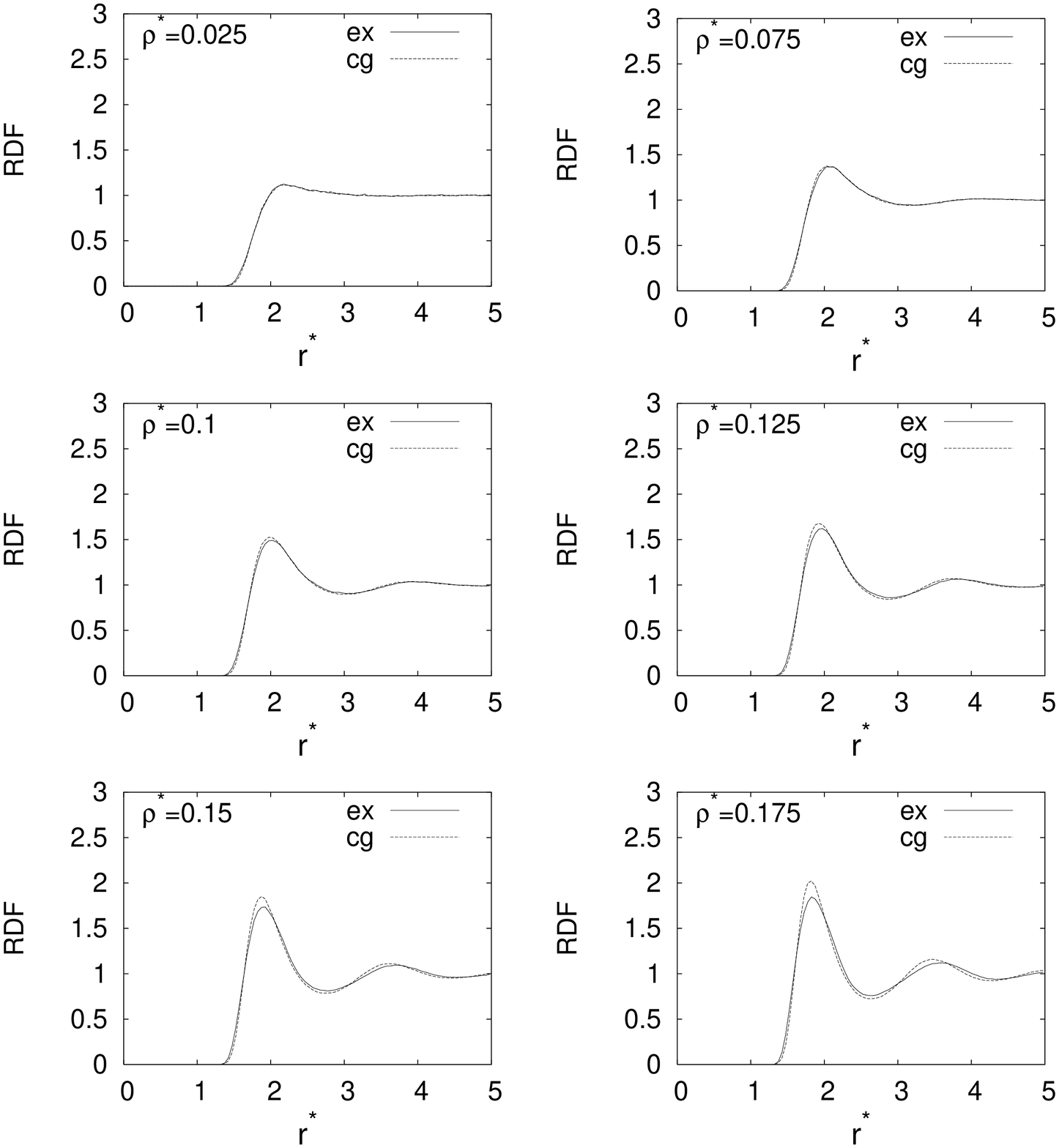}
\caption{Center-of-mass radial distribution functions of
the explicit (\emph{ex}) and coarse-grained (\emph{cg}) systems at
the temperature $T^*=1$ and number density $0.025\le\rho^*\le
2.25$.}\label{Fig.4}
\end{figure}

The obtained RDF$_{cm}$s of the \emph{ex} and \emph{cg} systems at
the temperature $T^*=1$ and number density $0.025\le\rho^*\le
0.175$ using for all number densities the same effective potential
given by Eq. (\ref{eq.11}) for the pair interactions between the
coarse-grained molecules are depicted in figure \ref{Fig.4}. 
The
RDF$_{cm}$s are calculated in the range $r^*\in [0,5]$ with
$\Delta r^*=0.05$. From results presented in figure \ref{Fig.4}
it can be observed that although the effective potential given by
Eq. (\ref{eq.11}) was parameterized at the number density
$\rho^*=0.1$  the agreement of RDF$_{cm}$s of the \emph{cg} system
with the corresponding reference RDF$_{cm}$s of the \emph{ex}
system is good also for the lower number densities. In fact, due
to weaker many-body interactions the agreement is even better for
the systems with lower density. Since the effective potentials are
density dependent\cite{Louis:2002}, to obtain a better agreement
for higher densities the effective potential should be
reparameterized\cite{Reith:2003}. Probably the functional form of
the effective potential will also change due to increased
contribution from many-body interactions.

As a quantitative measure of accuracy of the evaluated RDF$_{cm}$s
we define a penalty function $f_p$ as
\begin{equation}
 f_p=\int [g(r^*)-g^{cm}_{ex}(r^*)]^2\exp(-r^*)\,dr^*,\label{eq.10}
\end{equation}
where $g^{cm}_{ex}$, which is taken as a reference, is the
RDF$_{cm}$ of the \emph{ex} system. The function $\exp(-r^*)$ is
employed to penalize more strongly deviations at small
distances\cite{Reith:2003}. The values of $f_p$ for \emph{cg}
systems with the number density $0.025\le\rho^*\le 0.175$ are
reported in table \ref{Tab.1}.
\begin{table}[ht]
   \centering
  \begin{tabular}{||c|c||}
    \hline\hline
      $\rho^*$  & $f_p\cdot 10^3$        \\
    \hline
      $0.025$     & $0.0401$  \\
      $0.075$     & $0.1101$  \\
      $0.1$       & $0.2067$  \\
      $0.125$     & $0.4361$  \\
      $0.150$     & $0.9257$  \\
      $0.175$     & $1.9025$  \\
    \hline\hline
   \end{tabular}
   \caption{Penalty function $f_p$ defined by Eq. (\ref{eq.10}) as a function of number density
   $\rho^*$ for RDF$_{cm}$s $g^{cm}_{cg}(r^*)$ of the coarse-grained
   systems in which particles are interacting via the effective
   potential ${U^{cm}}^*$ given by
   Eq. (\ref{eq.11}). The RDF$_{cm}$s $g^{cm}_{ex}(r^*)$ of all-atom systems at the corresponding $\rho^*$
   are taken for the reference RDF$_{cm}$s.} \label{Tab.1}
  \end{table}
 As expected the $f_p$ grows with the growing density of the system.

\begin{figure}[!ht]
\centering
\includegraphics[height=10cm,width=7.5cm,angle=0]{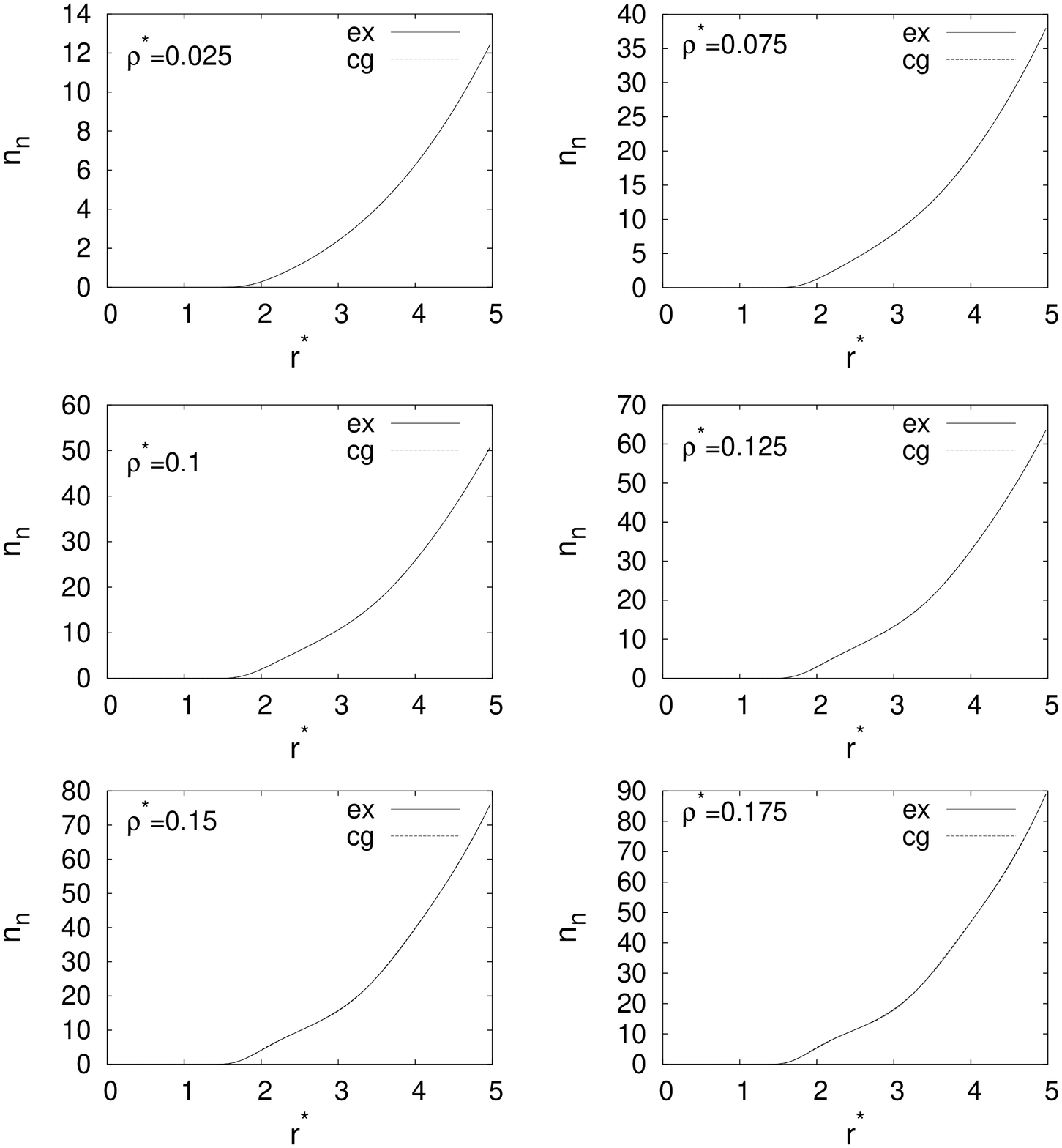}
\caption{The average number of neighbors $n_n(r^*)$ of a
given molecule as a function of distance for explicit (\emph{ex})
and coarse-grained (\emph{cg}) systems at the temperature $T^*=1$
and number density $0.025\le\rho^*\le 2.25$.}\label{Fig.5}
\end{figure}

From the RDF$_{cm}$ we can evaluate the average number of
neighbors of a given molecule within a sphere with the radius
$r^*$ as
\begin{equation}
  n_n(r^*)=\rho^*\int_0^{r^*} g(r^*)4\pi r^{*^2}\,dr^*.
\end{equation}
The $n_n(r^*)$ for the \emph{ex} and \emph{cg} systems at the
temperature $T^*=1$ and number density $0.025\le\rho^*\le 0.175$
are shown in figure \ref{Fig.5}. Despite the deviations between
the RDF$_{cm}$s of the \emph{ex} and \emph{cg} systems the
corresponding average numbers of neighbors exactly match,
indicating, together with the RDF$_{cm}$s presented in figure
\ref{Fig.4}, that the \emph{cg} model with the molecules
interacting via the effective potential given by Eq. (\ref{eq.11})
reproduces well the structure of the underlying \emph{ex} system.

\begin{figure}[!ht]
\centering
\mbox{\includegraphics[width=7.5cm, angle=0]{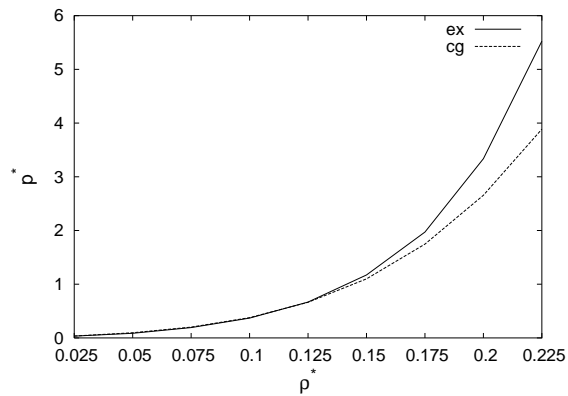}}
         \caption{Pressure $p^*$ in the explicit (\emph{ex}) and
coarse-grained (\emph{cg}) systems at the temperature $T^*=1$ as a
function of the number density $\rho^*$ of the system.}\label{Fig.6}
\end{figure}

To show that the \emph{cg} system with the effective potential
from Eq. (\ref{eq.11}) is at the same state point as the original
\emph{ex} system at the same temperature and density we also
evaluated the pressure in the system. The equations of state for
the \emph{ex} and \emph{cg} models are shown in figure
\ref{Fig.6}. 
The resulting equations of state are similar to the
case of colloidal sphere systems\cite{Bryant:2002}. This
similarity is to be expected since the tetrahedral molecule as
well as the spherical coarse-grained molecule can be considered as
spherical colloidal particles with a hard core and a soft coating
layer. From figure \ref{Fig.6} we can conclude that the \emph{cg}
model with the particles interacting via the effective potential
from Eq. (\ref{eq.11}) reproduces the equation of state of the
underlying \emph{ex} system up to $\rho^*=0.125$, at which the two
pressure curves start deviating from each other, reflecting the
fact that the effective potential is density dependent. In order
to minimize the artifacts originating from our parameterization of
the effective potential while still simulating the liquid state we
have decided to perform all our MD simulations of the hybrid
atomistic/mesoscopic (\emph{ex-cg}) model at the state point with
$T^*=1$ and density $\rho^*=0.1$.

\subsection*{Statistical Properties}

The new adaptive resolution scheme is tested by comparing the
computed statistical properties of the \emph{ex-cg} model with the
corresponding properties of the reference fully atomistic \emph{ex}
system.

\begin{figure}[!ht]
\centering
\subfigure[]{\includegraphics[width=7.5cm]{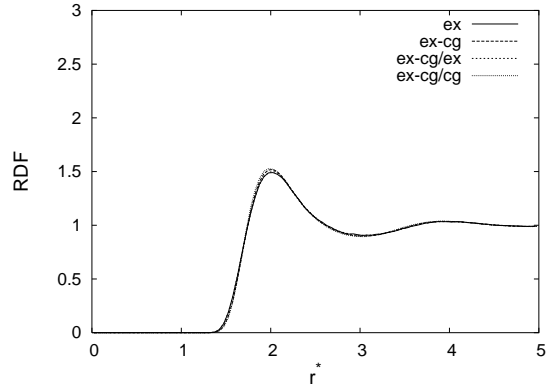}}\\
\subfigure[]{\includegraphics[width=7.5cm]{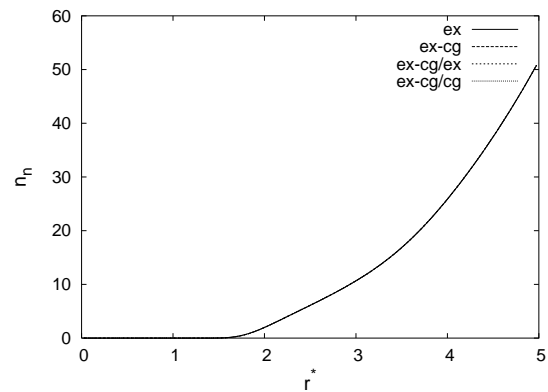}}
         \caption{(a) Center-of-mass radial distribution functions
for all molecules in the box of the all-atom (\emph{ex}) and hybrid
atomistic/mesoscopic (\emph{ex-cg}) systems at $\rho^*=0.1$ and
$T^*=1$.  Shown are also the corresponding center-of-mass radial
distribution functions for only the explicit molecules from the
explicit region (\emph{ex-cg}/\emph{ex}) and for only the
coarse-grained molecules from the coarse-grained region
(\emph{ex-cg}/\emph{cg}). The width of the interface layer is
$2d^*=2.5$. (b) The corresponding average numbers of neighbors
$n_n(r^*)$ of a given molecule as functions of distance. The
different curves are almost indistinguishable.}\label{Fig.7}
\end{figure}

Figure \ref{Fig.7} (a) displays the RDF$_{cm}$s calculated from
center-of-mass positions of all molecules in the box of the
\emph{ex} and \emph{ex-cg} systems at $\rho^*=0.1$ and $T^*=1$.
Shown is the case with the width of the interface layer
$2d^*=2.5$. Depicted are also the corresponding local RDF$_{cm}$s
for the explicit  (\emph{ex-cg}/\emph{ex}) and coarse-grained
regions (\emph{ex-cg}/\emph{cg}) of the \emph{ex-cg} model. As in
all simple liquids, e.g., methane, the RDF$_{cm}$s are zero at
short distances between molecules' centers of mass because
repulsive forces prevent overlapping of molecules. Then the
functions increase rapidly to the first peak. With increasing
distance RDF$_{cm}$s reach the limiting value $1$ after few
oscillations, indicating that there is no order at long distances.
The average numbers of neighbors $n_n(r^*)$ of a given molecule as
functions of distance are illustrated in figure \ref{Fig.7} (b).
The number of nearest neighbors in the first layer corresponding
to the first minimum in the RDF$_{cm}$ is about $11$. For comparison, the corresponding
experimental value from X-ray diffraction for liquid methane at
$T=92\units{K}$ is approximately $12$\cite{Habenschuss:1981}.

All calculated RDF$_{cm}$s of the \emph{ex-cg} model and
$n_n(r^*)$  are in good agreement with the reference RDF$_{cm}$
and $n_n(r^*)$ of the \emph{ex} model indicating that the
structure of the underlying all-atom system is well reproduced
using the adaptive resolution scheme. This is further confirmed by
the values of the penalty function $f_p$ reported in table
\ref{Tab.2}, evaluated for different interface layer widths. 
\begin{table}[h]
   \centering
   \begin{tabular}{||c|c|c|c||}
    \hline\hline
    \vbox to0pt{\hbox{\lower 0.6\baselineskip\hbox{$2d^*$}}\vss}&
    \multicolumn{3}{c||}{$f_p\cdot 10^3$}\\
    \cline{2-4}
           & ex-cg     & ex-cg/ex   & ex-cg/cg  \\
   \hline
    $2.5$  & 0.0821 &  0.0009 & 0.2109 \\
    $3.0$  & 0.0893 &  0.0008 & 0.2158 \\
    $4.0$  & 0.1058 &  0.0016 & 0.2197 \\
    $5.0$  & 0.1324 &  0.0023 & 0.2315 \\
    $6.0$  & 0.1583 &  0.0033 & 0.2367 \\
    $7.0$  & 0.1944 &  0.0047 & 0.2449 \\
    $8.0$  & 0.2301 &  0.0076 & 0.2616 \\
    $9.0$  & 0.2695 &  0.0082 & 0.2489 \\
   $10.0$  & 0.3159 &  0.0101 & 0.2632 \\
    \hline

    \hline\hline
   \end{tabular}
   \caption{Penalty function $f_p$ defined by Eq. (\ref{eq.10}) as a
   function of the interface layer width $2d^*$ for RDF$_{cm}$s
   $g^{cm}_{ex-cg}(r^*)$, $g^{cm}_{ex-cg/ex}(r^*)$,
   $g^{cm}_{ex-cg/cg}(r^*)$ of the hybrid atomistic/mesoscopic model
   at $\rho^*=0.1$ and $T^*=1$.  $g^{cm}_{ex-cg}(r^*)$ is the
   RDF$_{cm}$ of all molecules in the box where all molecules are
   considered indistinguishable, $g^{cm}_{ex-cg/ex}(r^*)$ is the RDF$_{cm}$ of only
   the explicit molecules from the explicit region while
   $g^{cm}_{ex-cg/cg}(r^*)$ is the RDF$_{cm}$ of only
   the coarse-grained molecules from the coarse-grained region. The RDF$_{cm}$ $g^{cm}_{ex}(r^*)$
   of all-atom system at the corresponding $\rho^*$ and $T^*$
   is taken for the reference RDF$_{cm}$.}\label{Tab.2}
\end{table}
From
the results given in table \ref{Tab.2} we can see that the local
structure in the explicit region of the \emph{ex-cg} model is
exactly reproduced independently of the interface layer width. In
contrast, the accuracy of the local structure reproduction in the
coarse-grained region of the \emph{ex-cg} model depends on the
accuracy of the effective potential parameterization in the
\emph{cg} model (see the value of $f_p$ given in table \ref{Tab.1}
for $\rho^*=0.1$). The computed $f_p$s for the total RDF$_{cm}$ of
the \emph{ex-cg} model lie up to $2d^*=8.0$ in between the values
for the local explicit and coarse-grained RDF$_{cm}$s.

For narrow interface layers with $2d^*<2.5$ the adaptive
resolution scheme fails to work. The molecules are not given
enough space and time to adapt their orientation to their
environment and the system is not properly equilibrated in this
case. The value $2d^*=2.5$, for which the adaptive resolution
scheme gives the acceptable results,  can be rationalized by the
fact that the interface layer width should at least exceed the
maximal range of interaction, which is the range of the effective
potential, namely $2.31\sigma$.

To demonstrate that the thermodynamic properties are correctly
reproduced by the adaptive resolution scheme the temperature and
pressure of the system as a functions of the interface layer width
are given in tables \ref{Tab.3} and \ref{Tab.4}, respectively. 
The calculated temperature profile reported in table \ref{Tab.3} shows
that the system is at the right temperature and that all degrees
of freedom are properly equilibrated in accordance with the
equipartition principle regardless of the interface layer width.
The results given in table \ref{Tab.4} also show that the adaptive
resolution scheme succeeds in reproducing the pressure of the
underlying fully atomistic system.
\begin{table*}[ht]
   \centering
  \begin{tabular}{||c|c|c|c|c|c|c||}
    \hline\hline
      $2d^*$  & $T$           & $T_{ex}$       & $T_{cg}$ & $T_{int}$ & $T_{ex}^{all}$& $T_{int}^{all}$        \\
    \hline
      $0^{ex}$& $1.00\pm 0.01$&  $1.00\pm 0.01$&  $-$     & $-$ &$1.00\pm 0.01$   &       $-$                   \\
      $0^{cg}$&  $1.00\pm 0.01$         &   $-$       &   $1.00\pm 0.01$  &    $-$         &    $-$           &  $-$ \\
      $2.5$   &   $1.00\pm 0.01$         &  $1.00\pm 0.01$          &   $1.00\pm 0.02$        &  $1.00 \pm 0.03$            &   $1.00\pm 0.01$            & $1.00\pm 0.02$  \\
      $3.0$   & $1.00\pm 0.01$ & $1.00 \pm 0.02$ &   $1.00 \pm 0.02$ &   $1.00 \pm 0.03$   &  $1.00 \pm 0.01$ &  $1.00 \pm 0.02$   \\
      $4.0$   &  $1.00\pm 0.01$        &  $1.00\pm 0.01$         &  $1.00 \pm 0.02$        &    $1.00 \pm 0.03$           &    $1.00\pm 0.01$             & $1.00\pm 0.01$   \\
      $5.0$   &  $1.00\pm 0.01$        &   $1.00\pm 0.02$        &  $1.00\pm 0.02$        &    $1.00\pm 0.02$           &     $1.00\pm 0.01$          &  $1.00\pm 0.02$ \\
      $6.0$   &  $1.00\pm 0.01$        &  $1.00\pm 0.02$          &  $1.00\pm 0.02$        &    $1.00\pm 0.02$          &        $1.00\pm 0.02$        &   $1.00\pm 0.02$\\
      $7.0$   &    $1.00\pm 0.01$       &    $1.00\pm 0.02$        &   $1.00\pm 0.02$       &     $1.00\pm 0.02$        &    $1.00\pm 0.02$           &  $1.00\pm 0.01$\\
      $8.0$   &   $1.00\pm 0.01$        & $1.00\pm 0.02$          &   $1.00\pm 0.02$       & $1.00\pm 0.02$            &          $1.00\pm 0.02$     &  $1.00\pm 0.01$\\
      $9.0$   &  $1.00\pm 0.01$        &     $1.00\pm 0.03$      & $1.00\pm 0.03$         &     $1.00\pm 0.02$        &        $1.00\pm 0.01$       &   $1.00\pm 0.01$ \\
      $10.0$  &    $1.00\pm 0.01$      &  $1.00\pm 0.03$         &    $1.00\pm 0.02$        &  $1.00\pm 0.02$           &          $1.00\pm 0.01$        & $1.00\pm 0.02$  \\

    \hline\hline
   \end{tabular}
   \caption{Average temperature as a function of the interface
 layer width $2d^*$. $T$, $T_{ex}$, $T_{cg}$, and $T_{int}$ are the average temperatures
 of the total system, the explicit, coarse-grained, and interface
 layer regions, respectively, calculated by Eq. (\ref{eq.12}). $T_{ex}^{all}$ and $T_{int}^{all}$
 are the average temperatures of the explicit and interface layer regions,
 respectively, calculated from total velocities
 (translational+vibrational+rotational) of explicit atoms in
 molecules. $0^{ex}$ and $0^{cg}$ denote the all-atom and
 coarse-grained systems, respectively. } \label{Tab.3}
  \end{table*}
\begin{table}[ht]
   \centering
  \begin{tabular}{||c|c||}
    \hline\hline
      $2d^*$        & $p^*$        \\
    \hline
      $0^{ex}$      &  $0.379\pm 0.009$\\
      $0^{cg}$      &  $0.378\pm 0.004$\\
      $2.5$         &  $0.382\pm 0.007$\\
      $3.0$         &  $0.383\pm 0.006$\\
      $4.0$         &  $0.384\pm 0.006$\\
      $5.0$         &  $0.385\pm 0.007$\\
      $6.0$         &  $0.386\pm 0.006$\\
      $7.0$         &  $0.388\pm 0.004$\\
      $8.0$         &  $0.389\pm 0.006$\\
      $9.0$         &  $0.390\pm 0.005$\\
     $10.0$         &  $0.391\pm 0.006$\\
    \hline\hline
   \end{tabular}
   \caption{Average pressure calculated using Eq. (\ref{eq.12a}) as a function of the interface
 layer width $2d^*$. $0^{ex}$ and $0^{cg}$ denote the all-atom and
 coarse-grained systems, respectively. } \label{Tab.4}
  \end{table}

\begin{figure}[!ht]
\centering
\mbox{\includegraphics[width=7.5cm, angle=0]{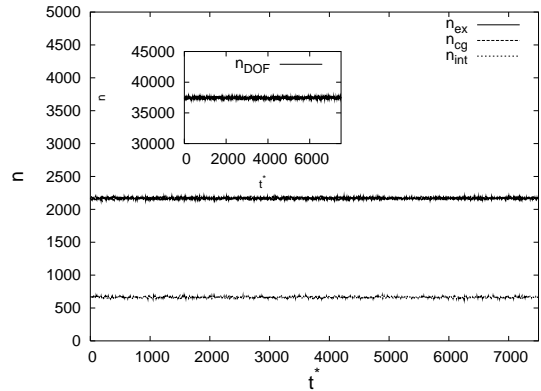}}
         \caption{ Time evolution of number of molecules in a
explicit ($n_{ex}$) , coarse-grained ($n_{cg}$), and interface
regions ($n_{int}$) in the hybrid atomistic/mesoscopic model with
the $2.5\sigma$ interface layer width. Time evolution of number of
degrees of freedom in the system ($n_{DOF}$) is depicted in the
inset.}\label{Fig.8}
\end{figure}

In order to check that the chemical potentials in the atomistic
and coarse-grained regions are equal as required by the condition
(\ref{eq.1}) we report in table \ref{Tab.5} the average number of
molecules in different regions of the system.  In table
\ref{Tab.5} we also give the number of degrees of freedom
$n_{DOF}$ in the system defined as
\begin{equation}
  n_{DOF}=3\sum_\alpha [w_\alpha N+(1-w_\alpha)],\label{eq.13}
\end{equation}
where $N=4$ is the number of explicit tetrahedral atoms in a
molecule and $w_\alpha$ is the value of the weighting function
defined by Eq. (\ref{eq.7}) for the molecule $\alpha$. The
summation in Eq. (\ref{eq.13}) goes over all $n$ molecules of the
system. 
\begin{table}[ht]
   \centering
  \begin{tabular}{||c|c|c|c|c||}
    \hline\hline
      $2d^*$  & $n_{ex}$ & $n_{cg}$ & $n_{int}$ & $n_{DOF}$       \\
    \hline
      $0^{ex}$ &  $5001$       &  $0$          &  $0$         &  $60012$\\
      $0^{cg}$ &    $0$        &  $5001$       &  $0$         &  $15003$\\
       $2.5$   &  $2167\pm 38$ &  $2170\pm 48$ & $663\pm 21$  &  $37446\pm 516$\\
       $3.0$   &  $2100\pm 45$ &  $2103\pm 60$ & $796\pm 37$  &  $37455\pm 493$\\
       $4.0$   &  $1966\pm 58$ &  $1969\pm 47$ & $1064\pm 40$ &  $37467\pm 507$              \\
       $5.0$   &  $1832\pm 47$ &  $1835\pm 56$ & $1332\pm 43$ &  $37470\pm 501$             \\
       $6.0$   &  $1697\pm 57$ &  $1701\pm 54$ & $1601\pm 34$ &  $37467\pm 330$             \\
       $7.0$   &  $1563\pm 45$ &  $1566\pm 29$ & $1871\pm 31$ &  $37479\pm 526$              \\
       $8.0$   &  $1428\pm 48$ &  $1431\pm 27$ & $2141\pm 22$ &  $37479\pm 345$              \\
       $9.0$   &  $1294\pm 22$ &  $1295\pm 23$ & $2411\pm 60$ &  $37488\pm 576$           \\
      $10.0$   &  $1158\pm 33$ &  $1160\pm 16$ & $2682\pm 56$ &  $37482\pm 63$              \\
    \hline\hline
   \end{tabular}
   \caption{Average number of molecules as a function of the interface
   layer width $2d^*$. $n_{ex}$, $n_{cg}$, and $n_{int}$ are the
   average number of molecules in the explicit, coarse-grained, and
   interface layer regions, respectively. $n_{DOF}$ is the average
   number of degrees of freedom defined by Eq. (\ref{eq.13}).
   For orientation: in the system with $2500$ coarse-grained molecules,
   $2500$ four atom explicit molecules, and no hybrid molecules
   $n_{DOF}=37500$. $0^{ex}$ and $0^{cg}$ denote the all-atom and
 coarse-grained systems, respectively.} \label{Tab.5}
  \end{table}
The results show that there is no molecule number bias in
the system and that the \emph{ex-cg} system is homogeneous. Note
that $n_{DOF}$ is greatly reduced employing the adaptive
resolution scheme on the \emph{ex-cg} model in comparison with the
fully atomistic model. The time evolution of the number of
molecules in different regions of the system together with the
$n_{DOF}$ is illustrated in figure \ref{Fig.8}.
 The results
clearly demonstrate that the system is in thermodynamical
equilibrium, indicating that the conditions (\ref{eq.1}) are
satisfied by our adaptive resolution scheme.

Although we have parameterized the effective potential to
reproduce the structural properties of the \emph{ex} system we can
also compare the dynamical properties of the \emph{ex-cg} with the
\emph{ex} model. For that purpose we have computed the
self-diffusion coefficient, which is evaluated from the
center-of-mass displacements using the Einstein relation
 \begin{equation}
  D^*=\frac{1}{6}\lim_{t^*\rightarrow\infty}\frac{\langle|{\bf
  R}^*_\alpha(t^*)-{\bf R}^*_\alpha(0)|^2\rangle}{t^*},\label{eq.14}
 \end{equation}
where ${\bf R}^*_\alpha(t^*)$ is the center-of-mass position of
the molecule $\alpha$ at time $t^*$ and averaging is performed
over all molecules and all choices of time origin. The
self-diffusion coefficient of the \emph{ex} and \emph{cg} models
calculated in the microcanonical ensemble, where the Langevin
thermostat is switched off after the initial warm-up run, are
$0.24$ and $0.30$, respectively. The corresponding values of the
self-diffusion coefficient for the \emph{ex}, \emph{cg}, and
\emph{ex-cg} models with the Langevin thermostat switched on are
$0.12$, $0.14$, and $0.13$, respectively. Since
we use the same time and length scales in all our three models
different values of the self-diffusion coefficient in the
\emph{ex} and \emph{cg} models (with no Langevin thermostat
applied) indicate that the coarse-grained molecules experience a
slightly smaller intermolecular frictional hindrance in their
motion compared to the explicit molecules. This indicates that the
effective pair potential given in Eq. (\ref{eq.11}) introduces an
effective time scale shift in the coarse-grained regime. This is
known from other studies \cite{Tschop:1998,Tschop:1998:2}, where
one actually takes advantage of that in order to reach very long
simulation times\cite{leon:2005}. The apparent self-diffusion
coefficient values of the \emph{ex} and \emph{cg} models are lower
and much closer together when the Langevin thermostat is applied
due to the frictional forces arising from the coupling to the
thermostat unlike to the case of polymers. There typically the
friction of the thermostat is negligible compared to the friction
between monomers. Since the self-diffusion coefficient of the
\emph{ex-cg} model is close to the corresponding values for the
\emph{ex} and \emph{cg} models we can conclude that the
center-of-mass dynamics of the molecules is similar in all three
models.

As the final test to demonstrate the reliability of the adaptive
resolution scheme we have computed the number density profile of
the \emph{ex-cg} model. 
\begin{figure}[!ht]
\centering
\subfigure[]{\includegraphics[width=7.5cm]{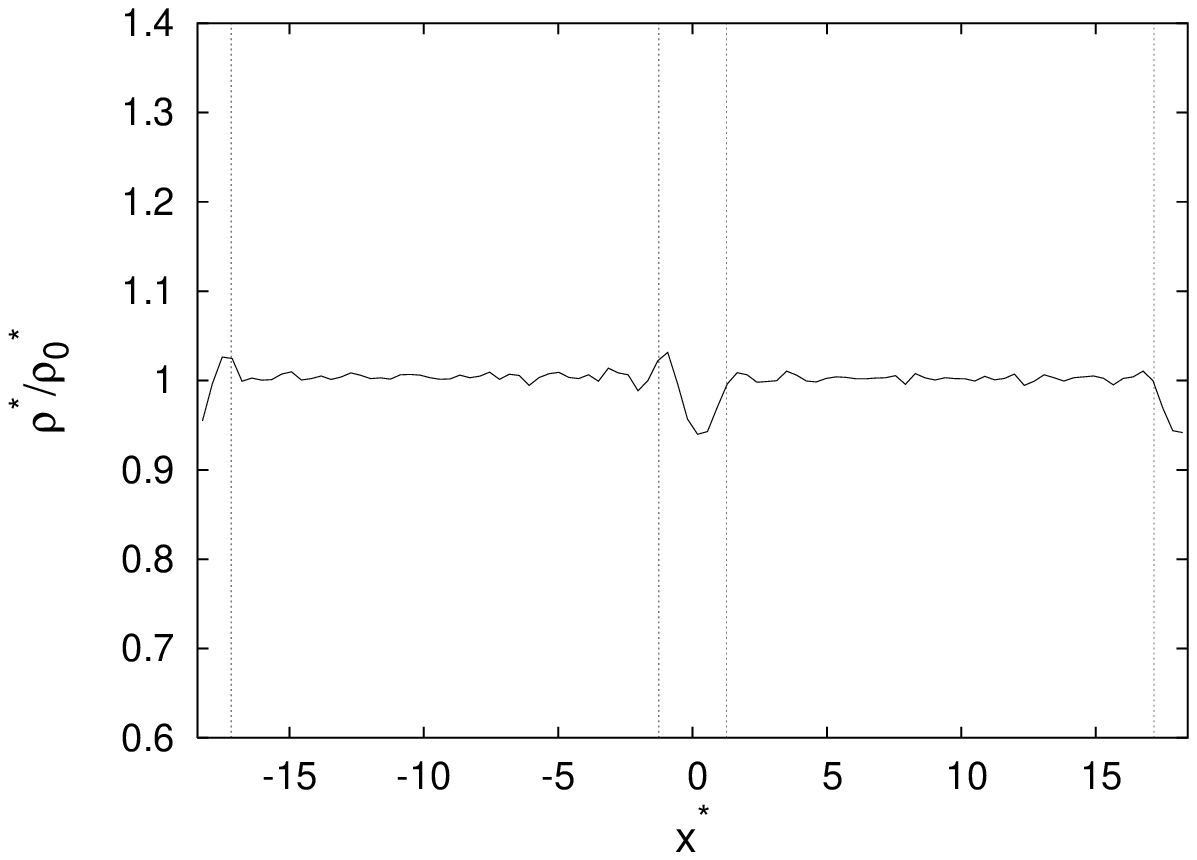}}\\
\subfigure[]{\includegraphics[width=7.5cm]{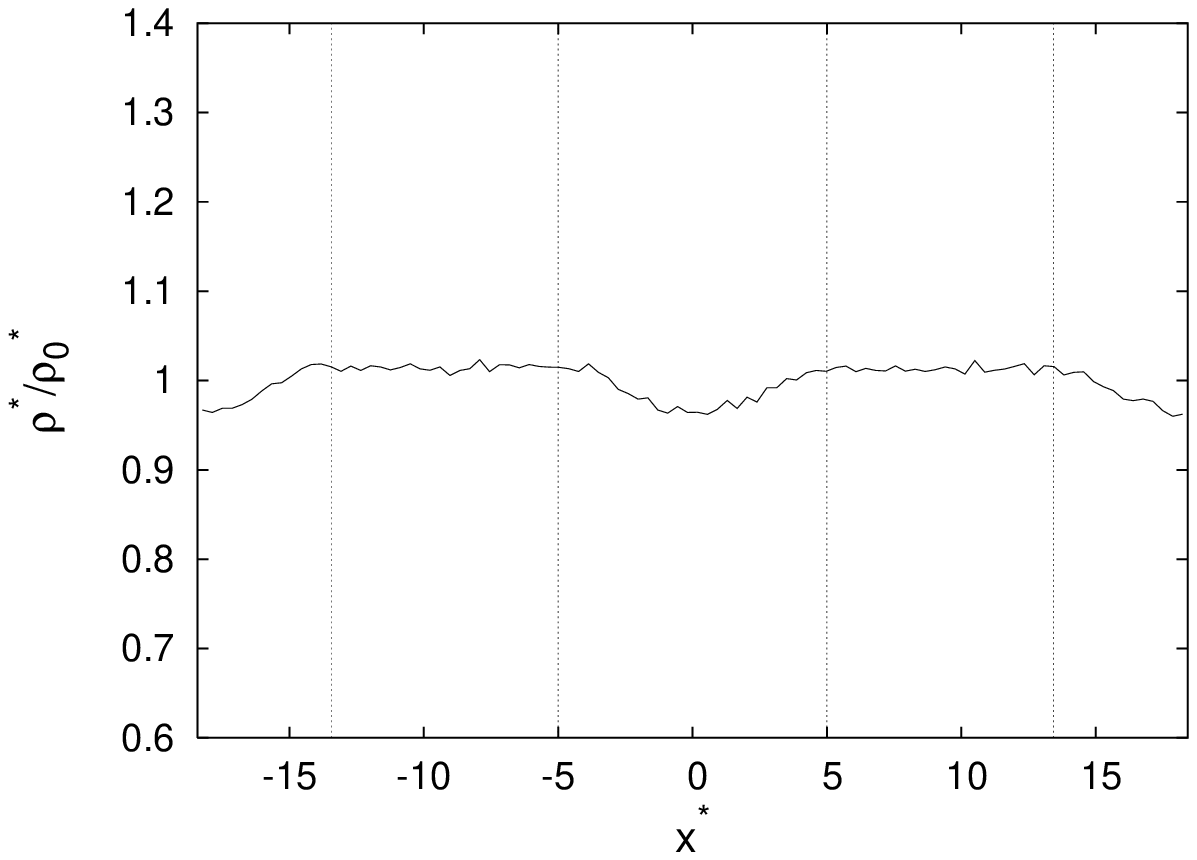}}

         \caption{(a) Normalized density profile in the $x$
direction of the hybrid atomistic/meso\-scopic model with the
$2.5\sigma$ interface layer width. Vertical lines denote
boundaries between atomistic, coarse-grained and interface regions
of the system. (b) The same as in (a) but for the  $10.0\sigma$
interface layer width.}\label{Fig.9}
\end{figure}
The results for the system with $2d^*=2.5$ and
$2d^*=10.0$ are presented in figures \ref{Fig.9} (a) and (b),
respectively. 
The results in figure \ref{Fig.9} (a) for the case
of $2d^*=2.5$ show that the explicit and coarse-grained regions
have the same homogeneous density as the reference system while
the density in the transition regime undergoes an oscillation
around the reference value $\rho^*_0=0.1$ with a magnitude of
approximately $0.05\rho^*_0$. In the case of $2d^*=10.0$ (figure
\ref{Fig.9} (b)) a $5\%$ drop in the density occurs in the
transition regime, which is compensated by the slight increase of
the density in the explicit and coarse-grained regions.

\begin{figure}[!ht]
\centering
\subfigure[]{\includegraphics[width=7.5cm]{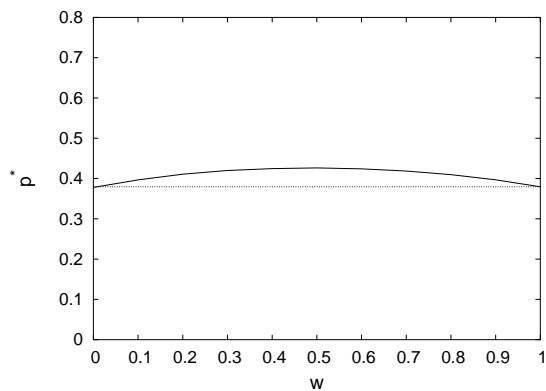}}\\
\subfigure[]{\includegraphics[width=7.5cm]{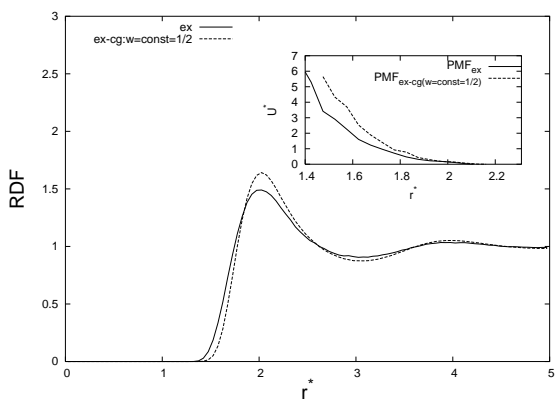}}
         \caption{Artifacts of the adaptive resolution scheme. (a)
Average pressure $p^*$ in the system containing only hybrid
molecules as a function of the constant value of the weighting
function $w$. (b) Center-of-mass radial distribution functions for
the explicit system (ex) and the system containing only hybrid
molecules with $w=const=1/2$ (\emph{ex-cg}(w=const=1/2)) at
$\rho^*=0.0025$ and $T^*=1$. The inset also shows the
corresponding PMF$_{ex}$ and PMF$_{ex-cg(w=const=1/2)}$ determined
from the systems with $\rho^*=0.0025$  using Eq. (\ref{eq.0}).
}\label{Fig.10}
\end{figure}

This artifact can be explained by considering the results
displayed in figure \ref{Fig.10} (a), where the pressure of the
system containing only hybrid molecules as a function of the
constant value of the weighting function $w$ (corresponding to the
situation in the interface layer) is illustrated.
 The pressure is
increased in comparison to the pressure in the reference system.
Clearly the increase is most prominent for the most 'artificial'
case with $w=1/2$, indicating that there still is a small
'pressure barrier' in the interface region causing the density
dip. This is also evident from results in figure  \ref{Fig.10}
(b), where the RDF$_{cm}$ and the potential of mean force of the
system with $w=1/2$ are shown. The effective potential in a system
containing only hybrid molecules with constant value of the
weighting function  changes in comparison to the all-atom system.
This means that the hybrid molecules in the interface layers of
the \emph{ex-cg} model experience too strong effective interaction
leading to the pressure variations in the interface layer. Since the
artifact occurs at constant values of $w$ it is an artifact of the
linear combination of forces in Eq. (\ref{eq.4}) and not of the
functional form of the weighting function $w$. It must be
emphasized, however, that this artifact of the proposed adaptive
resolution scheme is within a $5\%$ error and that similar
artifacts, occurring at the boundary of two domains with different
level of detail, are also characteristic for other hybrid
schemes\cite{Cai:2000}.

\begin{figure}[!ht]
\centering
\subfigure[]{\includegraphics[width=7.5cm]{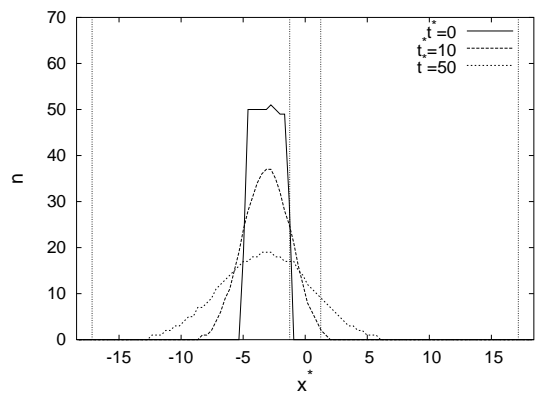}}\\
\subfigure[]{\includegraphics[width=7.5cm]{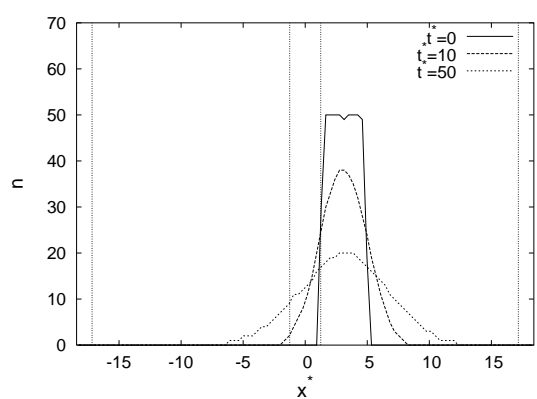}}
         \caption{Time evolution of diffusion profiles for the
molecules that are initially, at time $t^*=0$, localized at two
neighboring slabs of the mid interface layer with $2d^*=2.5$ ($n$
is the number of this molecules with the center-of-mass position
at a given coordinate $x^*$). The width of the two slabs is
$a^*/10$. Vertical lines denote boundaries of the interface layer.
(a) The diffusion profile, averaged over $500$ different time
origins, at $t^*=0$, $t^*=10$, and $t^*=50$ for the molecules that
are initially localized at the slab on the coarse-grained side of
the interface region. (b) The same as in (a) but for the molecules
that are initially localized at the slab on the atomistic side of
the interface region.
}\label{Fig.11}
\end{figure}

The pressure variations in the interface layer could cause a
spurious reflection of molecules from the boundary. However, the
results presented in figure \ref{Fig.11}, where the time evolution
of two diffusion profiles is monitored for molecules that are
initially localized at the two slabs with $a^*/10$ width
neighboring the interface layer, show that this is not the case.
The molecules initially localized at the two slabs spread out
symmetrically with time. This is because the self-diffusion
coefficients of all models are approximately the same in the case
of the applied Langevin thermostat. Thus, the two distributions
occupy at time $t^*$ regions with mean square radius
\begin{equation}
 \langle |x^*(t^*)-x^*(0)|^2\rangle\simeq 2D^*t^*,
\end{equation}
where $x^*(0)$ is the center of the distribution at time $t^*=0$,
which is equal to $-d^*-a^*/20$  for the left slab (figure
\ref{Fig.11} (a)) and $d^*+a^*/20$ for the right slab (figure
\ref{Fig.11} (b)). Since the diffusion profiles are symmetrical at
any given time we can conclude that the artifact at the interface
layer is too small to have any severe effect on the diffusion of
molecules across the interface layer.

\section{Conclusions}
A novel approach for efficient hybrid atomistic/mesoscale
molecular dynamics (MD) simulations has been presented in this
paper. The new adaptive resolution MD simulation scheme
dynamically couples the atomic and mesoscale length scales of the
studied system by allowing an on-the-fly dynamical interchange
between molecules' atomic and mesoscopic levels of description. In
our approach the number of degrees of freedom is allowed to
fluctuate during the course of simulation while the statistical
properties of the underlying all-atom model are properly
reproduced for both levels of detail. Since the purpose of this
paper was to develop the method, we restricted ourselves to a most
simple model system of a molecular liquid with short-range
repulsive interactions. Even on this level a number of questions
still have to be tackled, as there is the application to higher
liquid densities for complex molecular liquids or the proper
analysis and understanding of the the time scale problem, when the
diffusion on the coarse-grained level is faster than on the
all-atom level. For separate runs at the two levels of detail this
is understood, while the occurrence within one simulation box
still poses some conceptual problems. On the other hand the
current density is in between a typical small molecule liquid and
a polymeric fluid. Thus we expect that already this approach can
be generalized and applied to different realistic soft condensed
matter systems where both atomic and mesoscopic length scales have
to be considered. This can be either polymer solutions and melts,
such as a synthetic or biological macromolecule embedded in a
solvent. Similarly our method should also find application for
other polymer systems (same force field), molecular liquids such
as methane (same geometry) or water (tetrahedral clusters), etc.,
enabling to reach much larger length and time scales than for
all-atom MD simulations. In all cases the aim is to treat in a
simulation only as many degrees of freedom as absolutely necessary
for the question considered. In this sense the region of higher
detail can be either given by a geometrical constraint, e.g., close
to a surface, or even be chosen on demand due to specific local
conformations of a (macro-)molecular system. Work along these
lines is underway.
\vspace{0.5cm}

\section*{\small ACKNOWLEDGMENTS}
We thank A. Arnold, B.~A. Mann, P. Schravendijk, B. Hess, and N.
van der Vegt for useful discussions. We are also grateful to C.~F.
Abrams for discussions at early stage of this work. This work is
supported in part by the Volkswagen foundation. M.~P. acknowledges
the support of the Ministry of Higher Education, Science and
Technology of Slovenia under grant No. P1-0002.



\begin{thebibliography}{10}

\bibitem{Deuflhard:1999}
P.~Deuflhard, J.~Hermans, B.~Leimkuhler, A.~Mark, S.~Reich and R.~Skeel, editors,
\newblock {\em Computational Molecular Dynamics: Challenges, Methods, Ideas},
  volume~4 of {\em Lecture Notes in Computational Science and Engineering},
\newblock Springer Verlag, 1999.

\bibitem{Minary:2004}
P.~Minary, M.~E. Tuckerman, and G.~J. Martyna,
\newblock Phys. Rev. Lett. {\bf 93}, 150201 (2004).

\bibitem{Praprotnik:2005}
D.~Jane\v{z}i\v{c}, M.~Praprotnik, and F.~Merzel,
\newblock J. Chem. Phys. {\bf 122}, 174101 (2005).

\bibitem{Praprotnik:2005:1}
M.~Praprotnik and D.~Jane\v{z}i\v{c},
\newblock J. Chem. Phys. {\bf 122}, 174102 (2005).

\bibitem{Praprotnik:2005:2}
M.~Praprotnik and D.~Jane\v{z}i\v{c},
\newblock J. Chem. Phys. {\bf 122}, 174103 (2005).

\bibitem{Praprotnik:2004}
M.~Praprotnik, D.~Jane\v{z}i\v{c}, and J.~Mavri,
\newblock J. Phys. Chem. A {\bf 108}, 11056 (2004).

\bibitem{kremer:2001}
K.~Kremer, and F.~M\"uller-Plathe,
\newblock MRS Bulletin \textbf{26}, 205 (2001).

\bibitem{kremer:2004}
K.~Kremer in "Multiscale  Modelling and Simulation, Lecture Notes
on Computational Science and Engineering, S. Attinger and P.
Koumoutsakis eds, Springer 2004.

\bibitem{Ayton:2004}
G.~S. Ayton, H.~L. Tepper, D.~T. Mirijanian, and G.~A. Voth,
\newblock J. Chem. Phys. {\bf 120}, 4074 (2004).

\bibitem{Kranenburg:2003}
M.~Kranenburg, M.~Venturoli, and B.~Smit,
\newblock J. Phys. Chem. B {\bf 107}, 11491 (2003).

\bibitem{Nielsen:2004}
S.~O. Nielsen, C.~F. Lopez, G.~Srinivas, and M.~L. Klein,
\newblock J. Phys.: Condens. Matter {\bf 16}, R481 (2004).

\bibitem{Marrink:2004}
S.~J. Marrink, A.~H. de~Vries, and A.~E. Mark,
\newblock J. Phys. Chem. B {\bf 108}, 750 (2004).

\bibitem{Chang:2005}
R.~Chang, G.~S. Ayton, and G.~A. Voth,
\newblock J. Chem. Phys. {\bf 122}, 244716 (2005).

\bibitem{Cook:2005}
I.~R. Cooke, K.~Kremer, and M.~Deserno,
\newblock Phys. Rev. E {\bf 72}, 011506 (2005). 

\bibitem{Kremer:1990}
K.~Kremer and G.~S. Grest,
\newblock J. Chem. Phys. {\bf 92}, 5057 (1990).

\bibitem{Tschop:1998}
W.~Tsch{\" o}p, K.~Kremer, J.~Batoulis, T.~B{\" u}rger, and O.~Hahn,
\newblock Acta Polym. {\bf 49}, 61 (1998).

\bibitem{Tschop:1998:2}
W.~Tsch{\" o}p, K.~Kremer, O.~Hahn, J.~Batoulis, and T.~B{\" u}rger,
\newblock Acta Polym. {\bf 49}, 75 (1998).

\bibitem{Everaers:2004}
R.~Everaers, S.~K.~Sukumaran, G.~S.~Grest, C.~Svaneborg, A.~Sivasubramanian, and K.~Kremer,
\newblock Science {\bf 303}, 823 (2004).

\bibitem{Chun:2000}
H.~M. Chun, C.~E. Padilla, D.~N. Chin, M. Watanabe, V.~I. Karlov,
H.~E. Alper, K. Soosaar, K.~B. Blair, O.~M. Becker, L.~S.~D. Caves,
R. Nagle, D.~N. Haneym, and B.~L. Farmer,
\newblock J. Comput. Chem. {\bf 21}, 159 (2000).

\bibitem{Malevanets:2000}
A.~Malevanets and R.~Kapral,
\newblock J. Chem. Phys. {\bf 112}, 7260 (2000).

\bibitem{Villa:2005}
E.~Villa, A.~Balaeff, and K.~Schulten,
\newblock PNAS {\bf 102}, 6783 (2005).

\bibitem{DelleSite:2002}
L.~Delle Site, C.~F. Abrams, A.~Alavi, and K.~Kremer,
\newblock Phys. Rev. Lett. {\bf 89}, 156103 (2002).

\bibitem{Abrams:2003}
C.~F. Abrams, L.~Delle Site, and K.~Kremer,
\newblock Phys. Rev. E {\bf 67}, 021807 (2003).

\bibitem{DelleSite:2004}
L.~Delle Site, S.~Leon, and K.~Kremer,
\newblock J. Am. Chem. Soc. {\bf 126}, 2944 (2004).
\bibitem{Laio:2002}
A.~Laio, J.~VandeVondele and U.~Roethlisberger,
\newblock J.Chem.Phys. {\bf 116}, 6941 (2002).
\bibitem{Rafii:1998}
H.~Rafii-Tabar, L.~Hua, and M.~Cross,
\newblock J. Phys.: Condens. Matter {\bf 10}, 2375 (1998).

\bibitem{Broughton:1999}
J.~Q. Broughton, F.~F. Abraham, N.~B. Bernstein, and E.~Kaxiras,
\newblock Phys. Rev. B {\bf 60}, 2391 (1999).

\bibitem{Smirnova:1999}
J.~A. Smirnova, L.~V. Zhigilei, and B.~J. Garrison,
\newblock Comp. Phys. Commun. {\bf 118}, 11 (1999).

\bibitem{Csanyi:2004}
G.~Csanyi, T.~Albaret, M.~C. Payne, and A.~D. Vita,
\newblock Phys. Rev. Lett. {\bf 93}, 175503 (2004).

\bibitem{Abrams:2004}
C.~F. Abrams,
\newblock Inhomogenous coarse-graining of polymers and polymer/metal
  interfaces,
\newblock in {\em Computational Soft Matter: {F}rom Synthetic Polymers to
  Proteins}, edited by N.~Attig, K.~Binder, H.~Grubm{\" u}ller, and K.~Kremer,
  volume~23 of {\em {NIC} Series}, pages 275--288, John von {N}eumann
  {I}nstitute for {C}omputing, 2004.

\bibitem{Louis:2002}
A.~A. Louis,
\newblock J. Phys.: Condens. Matter {\bf 14}, 9187 (2002).

\bibitem{Curtarolo:2002}
S.~Curtarolo and G.~Ceder,
\newblock Phys. Rev. Lett. {\bf 88}, 255504 (2002).

\bibitem{Klapp:2004}
S.~H.~L. Klapp, D.~J. Diestler, and M.~Schoen,
\newblock J. Phys.: Condens. Matter {\bf 16}, 7331 (2004).

\bibitem{Reith:2003}
D.~Reith, M.~P{\" u}tz, and F.~M{\" u}ller-Plathe,
\newblock J. Comput. Chem. {\bf 24}, 1624 (2003).

\bibitem{Bryant:2002}
G.~Bryant, S.~R. Williams, L. Qian, I.~K. Snook, E. Perez, and F. Pincet,
\newblock Phys. Rev. E {\bf 66}, 060501(R) (2002).

\bibitem{Lawrence:2003}
C.~P. Lawrence and J.~L. Skinner,
\newblock Chem. Phys. Lett. {\bf 372}, 842 (2003).

\bibitem{Zwanzig:1954}
R.~W. Zwanzig,
\newblock J. Chem. Phys. {\bf 22}, 1420 (1954).

\bibitem{Leach:2001}
A.~R. Leach,
\newblock {\em Molecular Modelling},
\newblock Pearson Education Limited, Harlow, 2 edition, 2001.

\bibitem{Schwabl:1995}
F.~Schwabl,
\newblock {\em Quantum Mechanics},
\newblock Springer-Verlag, Berlin Heidelberg, 2 edition, 1996.


\bibitem{Soddemann:2003}
T.~Soddemann, B.~D{\" u}nweg, and K. Kremer,
\newblock Phys. Rev. E {\bf 68}, 046702 (2003).

\bibitem{Allen:1987}
M.~P. Allen and D.~J. Tildesley,
\newblock {\em Computer Simulation of Liquids},
\newblock Clarendon Press, Oxford, 1987.

\bibitem{Berendsen:1984}
H.~Berendsen, J.~Postma, W.~V. Gunsteren, A.~D. Nola, and J.~Haak,
\newblock J. Chem. Phys. {\bf 81}, 3684 (1984).

\bibitem{Mann:2004}
B.~A. Mann, R.~Everaers, C.~Holm, and K.~Kremer,
\newblock Europhys. Lett. {\bf 67}, 786 (2004).

\bibitem{Duenweg:1991}
B.~D{\" u}nweg and W.~Paul,
\newblock Int. J. Mod. Phys. C {\bf 2}, 817 (1991).


\bibitem{Kremer:2005}
The physically correct way of course would be
to calculate the second virial coefficient. This here is not that
helpful, because the interaction potential is rather smooth. The
present measure is more appropriate, when the structure of the
liquids is compared in an NVT simulation, as in our case.

\bibitem{Espresso:2005}
http://www.espresso.mpg.de.

\bibitem{Habenschuss:1981}
A.~Habenschuss, E.~Johnson, and A.~H. Narten,
\newblock J. Chem. Phys. {\bf 74}, 5234 (1981).


\bibitem{leon:2005}
S.~Leon, N.~van der Vegt, L.~Delle Site, and K.~Kremer,
\newblock Macromolecules in press (2005).

\bibitem{Cai:2000}
W.~Cai, M.~de~Koning, V.~V. Bulatov, and S.~Yip,
\newblock Phys. Rev. Lett. {\bf 85}, 3213 (2000).



\end{thebibliography}
\end{document}